\documentclass[notitlepage,twocolumn,showkeys,nofootinbib]{revtex4-1}

\usepackage{amsmath}
\usepackage{amssymb}

\usepackage{graphicx}

\newcommand{\be}[0]{\begin{equation}}
\newcommand{\ee}[0]{\end{equation}}
\newcommand{\bea}[0]{\setlength\arraycolsep{2pt}\begin{eqnarray}}
\newcommand{\eea}[0]{\end{eqnarray}}
\newcommand{\ui}{\mathrm{i}}
\newcommand{\ud}{\mathrm{d}}

\newcommand{\Z}{{\cal Z}}

\newcommand{\mb}{\mathbf}
\newcommand{\rmi}{\mathrm{i}}

\allowdisplaybreaks

\begin{document}

\title{A ray-optical Poincar\'e sphere for structured Gaussian beams}

\author{Miguel A.~Alonso}

\affiliation{The Institute of Optics, University of Rochester, Rochester NY 14627, USA}
\affiliation{Center for Coherence and Quantum Optics, University of Rochester, Rochester NY 14627, USA}
\email{alonso@optics.rochester.edu}

\author{Mark R.~Dennis}

\affiliation{H.~H.~Wills Physics Laboratory, University of Bristol, Tyndall Avenue, Bristol BS8 1TL, UK}

\begin{abstract}
A general family of structured Gaussian beams naturally emerges from a consideration of families of rays.
These ray families, with the property that their transverse profile is invariant upon propagation (except for cycling of the rays and a global rescaling), have two parameters, the first giving a position on an ellipse naturally represented by a point on the Poincar\'e sphere (familiar from polarization optics), and the other determining the position of a curve traced out on this Poincar\'e sphere.
This construction naturally accounts for the familiar families of Gaussian beams, including Hermite-Gauss, Laguerre-Gauss and Generalized Hermite-Laguerre-Gauss beams, but is far more general.
The conformal mapping between a projection of the Poincar\'e sphere and the physical space of the transverse plane of a Gaussian beam naturally involves caustics.
In addition to providing new insight into the physics of propagating Gaussian beams, the ray-based approach allows effective approximation of the propagating amplitude without explicit diffraction calculations.
\end{abstract}


\maketitle

\section{Introduction}

Structured Gaussian beams are amongst the most familiar examples of paraxially propagating light beams.
These include the Hermite-Gaussian (HG) beams \cite{Kogelnik,SiegmanLasers}, with intensity patterns resembling Cartesian grids, and Laguerre-Gauss (LG) beams \cite{Allen}, whose intensities are concentric rings and whose phase carries orbital angular momentum (OAM).
A remarkable feature of Gaussian beams is that their intensity profile does not change on propagation, apart from an overall scaling; even in the far field, HG and LG modes appear the same.
More recently, other self-similar beams have been studied in detail, including Airy beams \cite{BerryBalazs,Christodoulides}, which are self-similar on propagation up to a parabolic lateral displacement, and Bessel beams \cite{Durnin,Eberly} and Mathieu beams \cite{Mathieu} whose intensity profile does not change at all on propagation.
Self similarity is more than a mathematical peculiarity, and is an important aspect of many applications of structured light. 
For example, given the constant width of their main intensity lobe, approximations to Airy and Bessel beams have been the basis of several imaging techniques, whether for illumination to increase axial resolution \cite{Betzig,Dholakia} or for 3D shaping of the point-spread function to increase depth of focus \cite{Dowski,Cathey,Botcherby,Lee,Fahrbach}. 
Given their characteristic intensity and phase profiles, structured Gaussian and other self-similar beams have been used extensively for particle manipulation \cite{ONeil,Baumgartl,Garces,Meyrath,Woerdemann}, and may be eigenfunctions of natural optical operators, such as Bessel beams and LG beams of azimuthal order $\ell$, which carry an OAM of $\ell \hbar$ per photon \cite{Allen}.

Here we describe an approach to structured Gaussian beams in terms of ray optics.
Geometric optics is usually applied in situations where a light field has well-defined extended wavefronts with slow amplitude variations. 
However, it turns out Gaussian beams and their generalizations are remarkably amenable to such an analysis.
As fundamental Gaussian beams, as well as HG and LG beams, are modes of laser cavities with curved mirrors \cite{Kogelnik,SiegmanLasers}, their dynamics is well-approximated by a two-dimensional isotropic harmonic oscillator representing the transverse plane, with the mirror curvature acting as the harmonic potential.
Classical orbits in the two-dimensional isotropic oscillator are of course ellipses, and can be represented by points on the Poincar\'e sphere, more familiar in representing the polarization of a harmonic electric field \cite{Poincare}.
In our analysis, a Gaussian beam is represented by a two-parameter family of rays; the rays are divided into subfamilies describing ellipses which propagate in a self-similar way, and which are described by points on a Poincar\'e-like sphere for rays.
The choice of the other parameter of the ray family then corresponds to determining a closed path of ray ellipses on this sphere, which is different for different types of beams.
Consistency with ray optics forces quantization conditions on these parameters, both around the ellipse and on the Poincar\'e sphere path.  These conditions give, for certain natural choices of path, the quantum numbers associated with HG and LG modes \cite{OtherPaper}, although many new kinds of structured Gaussian beam are possible. 
An immediate generalization is to the generalized Hermite-Laguerre-Gaussian beams (HLG) \cite{PadgettCourtial,Agarwal,Calvo,HabrakenSPIE,HabrakenOL,HabrakenJMP,MilionePoincare}, which interpolate between the HG and LG families on a generalized Poincar\'e sphere via an anamorphic fractional Fourier transformation, realized physically by transforming HG or LG beams through a beam shaping device consisting of suitably chosen pairs of cylindrical lenses \cite{AV,AV2,Beijersbergen}.

We therefore are discussing objects very familiar in modern paraxial optics: mode families, optical operators, geometric optics and Poincar\'e spheres, although combined in what we believe is a new way.
The approach describes the general behavior of the beams to the level of providing interpretations of the Gouy phase (whose significance has long been disputed) \cite{Gouy,Simon,Subbarao,Winful} and geometric (Pancharatnam-Berry) phase \cite{GalvezJOSAA,GalvezPRL,MilionePhase}, in a way that reveals the hidden geometry behind the transverse spatial structure of these familiar light beams. 
Furthermore, well-established methods of approximating the wave fields from the ray family are highly efficient for this approach, and even give the analytic forms for HG and LG beams. 
In a way, it forms a more complete and intuitive approach to our operator-based description of Gaussian beams in \cite{OtherPaper}.
Our emphasis throughout is in recasting known properties of Gaussian beam families in terms of rays; the methods can be readily adapted as a design tool for new kinds of structured light.

The structure of this paper proceeds as follows.
In the next section we discuss elliptic families of rays and show how they are associated with a Poincar\'e sphere.
This is followed in Section \ref{sec:quant} by a discussion on their quantization, and in Section \ref{Pdisks} by their geometric representation.
Families of these ellipses and their quantization are discussed in Section \ref{sec:fams}, which are combined to give a general method of constructing approximate wave solutions (Section \ref{sec:recon}), which are then applied to the HG, LG and GHL beams (Section \ref{sec:exs}).
Properties such as Gouy and geometric phases (Section \ref{GGP}) and the generalization to other beams families such as Bessel and Airy beams (Section \ref{sec:sepself}) follow, before a concluding discussion.
Many additional proofs and derivations are presented in the Appendices.

\section{Elliptic orbits and the Poincar\'e sphere} \label{sec:ell}

In geometric optics, the complex amplitude function representing a propagating, coherent monochromatic scalar light field is associated with a two-parameter family of light rays.
For fields with slowly-varying intensities such as plane or spherical waves, the rays are normal to the wavefronts and intensity is proportional to the ray density. 
However, for fields with more spatial structure, the ray-wave connection is more subtle, as several rays may pass through a given point, albeit with different directions.
In general, families of rays are bounded by envelopes known as caustics \cite{berryupstill,nye:natural}. 
Interference fringe-like structures can be caused by overlapping sets of rays, propagating in different directions.
Near caustics or other features of structured light, the rays can differ significantly from the wavefront normals, but there is still a tight link between the wave and ray descriptions, and it is possible to accurately reconstruct the wave field by associating a complex contribution to the rays \cite{Uniform}. 
We will see how the geometry of structured Gaussian beams can be readily understood using rays.

\begin{figure}[htbp]
\centering
\fbox{\includegraphics[width=\linewidth]{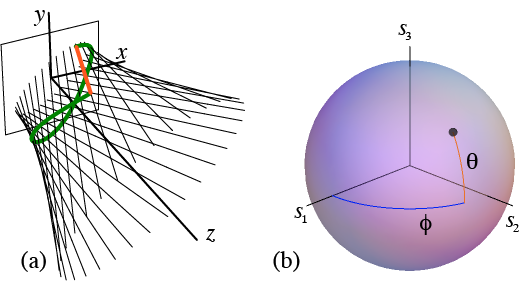}}
\caption{ Ray orbit in real space, and on the Poincar\'e sphere.
   (a) The straight rays sweep out a hyperboloid whose cross sections at any $z$ are ellipses of the same eccentricity and orientation. 
   The green curve is a normal to the rays. 
   The length of the orange ray segment must be an integer multiple of the wavelength. 
   (b) The eccentricity and orientation of the ellipse correspond to a point on the Poincar\'e sphere.
   } 
\label{Fig1} 
\end{figure}

We assume the beam propagates in a uniform, linear and isotropic medium so the rays are straight lines.
Each ray is determined by the transverse coordinate $\mb{Q} = (Q_x,Q_y)$ where it crosses the $z = 0$ plane and its direction $\mb{P} = (P_x,P_y)$.
The equation for the point where the ray crosses the plane of constant $z$ is therefore $\mb{Q} + z \mb{P}$.
For a beam to be self-similar on propagation, the arrangement of the rays should be the same (apart from overall scaling) as $z$ increases.
As in Figure \ref{Fig1}(a), both the shape and the orientation of the elliptic cross-section of a ruled hyperboloid are unchanged on propagation; endowing each ray in this one-dimensional subfamily with the same amplitude indeed guarantees its self-similarity on propagation.
The two-parameter family of rays making up structured Gaussian beams are therefore a one-parameter superfamily of elliptic families of rays.
We will call each such elliptic family an \emph{orbit} of rays.
In the quantification of these elliptic orbits, it is very convenient to use the parametrization of oriented ellipses afforded by the Poincar\'e sphere, borrowing language from polarization optics.

The Poincar\'e sphere for polarization parametrizes the two-dimensional complex Jones vectors $\mb{v}$ satisfying $\mb{v}^{\ast}\cdot\mb{v} = 1$, and $\mb{v}$ and $\mb{v}\exp(-\rmi \tau)$ are associated with the same polarization state for any real $\tau$.
$\mb{v}$ is defined in terms of latitude $\theta$ (Nb.~not colatitude) and azimuth $\phi$ on the unit Poincar\'e sphere, 
\begin{equation}
   \mb{v}(\theta,\phi)=\cos\frac{\theta}{2}\left(\cos\frac{\phi}{2},\sin\frac{\phi}{2}\right) + \rmi \sin\frac{\theta}{2}\left(-\sin\frac{\phi}{2},\cos\frac{\phi}{2}\right),
   \label{Jones}
\end{equation}
where $-\tfrac{1}{2}\pi \le \theta \le \tfrac{1}{2}\pi$ and $0 \le \phi < 2\pi$.
As $\tau$ varies, $0 \le \tau < 2\pi$,  $\Re[\mb{v}\exp(-\rmi \tau)]$ traces out the ellipse.
In polarization optics, $\mb{v}$ represents the transverse, harmonic electric field; with $\tau$ evolving as time, the real part gives the ellipse; for each $\tau$, the imaginary part is the velocity of the electric field vector \cite{OtherPaper}.

For the hyperboloidal orbit of rays, $\theta$ and $\phi$ are fixed parameters determining the eccentricity and orientation of the elliptic cross-section.
A given ray, labelled by $\tau$, has $z=0$ position
\begin{equation}
   \mb{Q}(\tau;\theta,\phi) = Q_0\,\Re\left[\mb{v}(\theta,\phi)\exp(-\rmi\tau)\right],
   \label{Qellips}
\end{equation}
where the constant $Q_0$ sets the transverse scale.
The ellipse's major and minor semi-axes are $Q_0\cos\tfrac{1}{2}\theta$ and $Q_0|\sin\tfrac{1}{2}\theta|$, and its foci are $\mb{f}_{\pm} = \pm Q_0 \cos^{1/2}\theta\,(\cos\tfrac{1}{2}\phi,\sin\tfrac{1}{2}\phi)$. 
The ray's direction---that is, its transverse velocity---is given by the imaginary part
\begin{equation}
   \mb{P}(\tau;\theta,\phi) = P_0\,\Im\left[\mb{v}(\theta,\phi)\exp(-\rmi\tau)\right],
   \label{Pellips}
\end{equation}
where $P_0$ is a constant determining the beam's angular divergence.
Thus, at any $z$, the transverse ray coordinates given by $\mb{Q}+z \mb{P}$ trace the same ellipse, up to a global hyperbolic scaling:
\begin{equation}
   \mb{Q} + z \mb{P} = \sqrt{Q_0^2+z^2P_0^2}\,\Re\left\{ \mb{v}(\theta,\phi)\exp[-\rmi(\tau+\zeta)]\right\},
   \label{rayGouy}
\end{equation}
where $\zeta=\arctan(zP_0/Q_0)$. 
On the ellipse, the position of each ray changes with $z$ (hence ``orbit''), but the orientation and eccentricity are unchanged, as shown in Fig.~\ref{Fig1}(a).

We stress that the parametrization of elliptic orbits of rays by a Poincar\'e sphere is different physically from polarization.
The similarity originates from the fact that mathematically, the Poincar\'e sphere parametrizes the classical orbits of the isotropic two-dimensional harmonic oscillator (like a transversely oscillating monochromatic electric field).
Less obviously, ray families propagating back-and-forth in laser cavities also behave like classical harmonic oscillators, as the curvature of the spherical mirrors effectively acts as an attractive harmonic potential for the rays. 
Structured Gaussian beams are made up of families of orbits described by paths on the Poincar\'e sphere.
First we discuss how the ray family is made compatible with the wave picture by a semiclassical `quantization condition'.

\section{Quantization condition for the orbits}\label{sec:quant}

Making the ray families consistent with wave optics requires two closure conditions dictated by the field's wavelength $\lambda$. 
The allowed solutions with certain properties (such as quantized OAM) are discrete, and often can be expressed as eigenfunctions of certain  operators.
These conditions are mathematically analogous to those in quantum mechanics, so we refer to them as {\it quantization} conditions. 
The first condition applies to the orbits. 
Since the rays in an orbit are skewed, a curve normal to them does not close onto itself after tracing the orbit (such as the thick green curve in Fig.~\ref{Fig1}(a)). 
There is a path difference along a ray between the initial and final points (represented by the orange line segment in Fig.~\ref{Fig1}(a)). Since optical path length (OPL) times wavenumber corresponds to the phase associated to a ray, this path difference must be an integer multiple of the wavelength for the ray and wave pictures to be consistent.

This condition is expressed mathematically as follows. 
Let $L_1(\tau)$ represent the OPL along each ray in the orbit, from some reference surface normal to the rays up to the $z=0$ plane. 
The rays' inclination is determined by ${\bf P}(\tau)$, so $L_1$ depends on $\tau$ as $\ud L_1={\bf P}\cdot\ud{\bf Q}$. 
From (\ref{Qellips}) and (\ref{Pellips}), the OPL difference between any pair of rays labelled by $\tau_1$ and $\tau_2$ is then
\begin{eqnarray}
   L_1(\tau_2)  & \!\!\!\!-\!\!\!\!& L_1(\tau_1) = \int_{\tau_1}^{\tau_2} {\bf P}\cdot\frac{\ud{\bf Q}}{\ud\tau}\,\ud\tau \nonumber \\
   & \!\!\!\!= \!\!\!\!& \frac{Q_0P_0}{2} \left[\tau_2-\tau_1-\frac{\sin(2\tau_2)-\sin(2\tau_1)}2\cos\theta\right]. \label{intL1}
\end{eqnarray}
After tracing the entire orbit, the total OPL mismatch is $L_1(2\pi)-L_1(0)=\pi Q_0P_0$, so the quantization condition yields
\begin{equation}
   Q_0P_0 = (N+1)\lambda/\pi,
   \label{quant1}
\end{equation}
where $N$ is a nonnegative integer. 
Significantly, this condition does not involve $\theta$ and $\phi$. 
Since $Q_0$ and $P_0$ describe the waist size and directional spread of the beam respectively, $Q_0P_0\pi/\lambda$ is the {\it beam quality factor} $M^2$ \cite{Sasnett,Siegman,Johnston}, usually defined as the ratio of the product of the spatial and directional widths of a beam to the same product for a fundamental Gaussian beam. 
Therefore, the beam quality factor of fields made up of orbits satisfying (\ref{quant1}) is quantized according to $M^2=N+1$. 
As we will discuss later, this index is also proportional to the beam's Gouy phase shift. 
In the quantum mechanical analogy, the integral (\ref{intL1}) plays the role of the semiclassical Bohr-Sommerfeld integral, whose quantization (\ref{quant1}) corresponds to energy quantization.
With light beams constructed from orbits satisfying (\ref{quant1}) with the same $N$ guarantees the profile will have a well-defined beam quality factor and Gouy phase, as the beam is constructed to be an eigenfunction of the corresponding Hamiltonian operator.

\section{Poincar\'e and physical disks}\label{Pdisks}

\begin{figure}[htbp]
\centering
\fbox{\includegraphics[width=\linewidth]{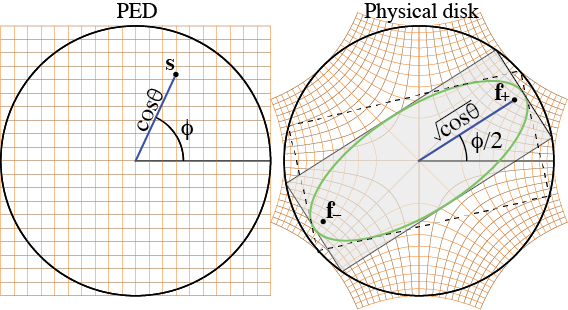}}
\caption{
   A point $\mb{s}$ in the PED maps to an ellipse with foci ${\bf f}_{\pm}$ in the physical disk. 
   The ellipse's major and minor axes have lengths $2\cos\tfrac{1}{2}\theta$ and $2|\sin\tfrac{1}{2}\theta|$ respectively, equal to the sides of the gray rectangle. 
   Note that any rectangle in which the ellipse is inscribed is itself inscribed in the unit circle.} 
\label{Fig2} 
\end{figure}

The shape of an elliptical orbit depends on $|\theta|$, with $\operatorname{sign} \theta$ determining the sense of twist of the rays around the ellipse under propagation in $z$ (that is, the sign of the OAM of the orbit $\mb{Q}\times\mb{P}$, which is positive (counterclockwise) in Figure \ref{Fig1}(a)). 
The beam's intensity profile is independent of this sign, so it is convenient to project both hemispheres of the Poincar\'e sphere, $\theta \gtrless 0$, onto the unit {\it Poincar\'e equatorial disk} (PED)\footnote{The term `Poincar\'e disk' is already used for a geometrical object \cite{Penrose}.}, with coordinates $\mb{s} = (s_1,s_2) = \cos\theta(\cos\phi,\sin\phi)$, and $|\mb{s}|^2 \le 1$. 
In the real space describing the transverse plane of the beam, we define the normalized ray position $\mb{q} = \mb{Q}/Q_0$.
The orbits are constrained to the interior of the unit disk $|\mb{q}|^2 \le 1$, which we call the {\it physical disk}, since it is a scaled version of a cross-section of the beam (for any $z$). 
As Fig.~\ref{Fig2} shows, a point ${\bf s}$ in the PED maps to an ellipse in the physical disk with foci $\mb{f}_{\pm}=\pm\sqrt{\cos\theta}\,(\cos\tfrac{1}{2}\phi,\sin\tfrac{1}{2}\phi)$, whose size is such that any rectangle in which it is inscribed is itself inscribed in the unit circle.

The mapping between the PED and physical disk can be appreciated mathematically by considering each as the unit disk in the complex plane, so any real vector $\mb{z}=(z_x,z_y)$ corresponds to the complex number $\Z(\mb{z}) = z_x + \rmi z_y = (1,\rmi)\cdot\mb{z}$. 
The complex numbers corresponding to the ellipse foci ${\bf f}_{\pm} = \pm\sqrt{\cos\theta}(\cos\tfrac{1}{2}\phi,\sin\frac{1}{2}\phi)$ are then the two square roots of the PED coordinate $\mb{s} = \cos\theta (\cos\phi,\sin\phi)$, as shown in Fig.~\ref{Fig2},
\begin{equation}
   \Z({\bf f}_{\pm})=\pm\sqrt{\Z({\bf s})}.
   \label{Zmap}
\end{equation}
This map is conformal (angle preserving) except at the origin, as shown in Fig.~\ref{Fig2}: a Cartesian grid over the PED maps onto a curvilinear orthogonal grid over the physical disk.

\section{Families of orbits, caustics and solid angle quantization condition}\label{sec:fams}

\begin{figure}[thbp]
\centering
\fbox{\includegraphics[width=\linewidth]{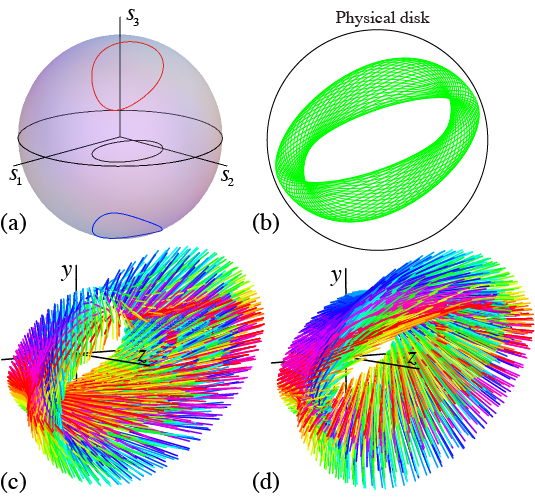}}
\caption{ 
   Ray families for a Poincar\'e path.
   (a) Two Poincar\'e paths (red and blue) over the surface of the Poincar\'e sphere with the same projection (black) onto the equatorial plane (the PED). 
   (b) Family of elliptical orbits for the Poincar\'e path in (a). 
   The inner and outer envelopes of this family form caustics. 
   (c,d) Rays corresponding to the loops in the (c) upper and (d) lower hemispheres, where colors identify orbits.} 
\label{Fig4} 
\end{figure} 

The complete two-parameter ray family is constructed as a continuous one-parameter set of orbits. 
For the global ray structure to be preserved on propagation, all orbits must be coaxial, share a waist plane, and have common $Q_0$ and $P_0$ (and hence $N$), so that they all scale as $(Q_0^2+z^2P_0^2)^{1/2}$. 
Such a set of orbits corresponds to a path on the Poincar\'e sphere, which we call a {\it Poincar\'e path}. 
The cases of interest here are those whosepaths are closed loops. 
For simplicity, consider first a Poincar\'e path confined to a hemisphere, so that its projection onto the PED is a closed loop that does not touch the disk's edge. 
There are two Poincar\'e paths (one in the upper hemisphere, one in the lower) projecting to each such PED loop, as shown in Fig.~\ref{Fig4}(a). 
Each point on the projected Poincar\'e path corresponds to an ellipse in the physical disk, so the complete closed path gives rise to a family of ellipses, as shown in Fig.~\ref{Fig4}(b).
Figs.~\ref{Fig4}(c,d) show how the shape of their transverse ray structures is preserved under propagation (up to a hyperbolic scaling), and that this structure is the same when the loop is in the (c) upper or (d) lower hemisphere; the hemisphere only determines the handedness (the sign of the OAM). 

Fig.~\ref{Fig4}(b) represents the beam as a superposition of elliptical ray orbits. 
This structure is determined by the path's projection onto the PED, which we will also refer to as the Poincar\'e path. 
The envelopes of the family are caustics, here an outer one enclosing all the rays, and an inner one inside of which there are no rays. 
The brightest intensity features of a beam are associated with these caustics, as the density of rays is highest near them.
There are surprisingly simple geometrical relations between the projected Poincar\'e path and the caustics in the physical disk, which we now describe (the derivation can be found in Appendix A).

\begin{figure}[htbp]
\centering
\fbox{\includegraphics[width=\linewidth]{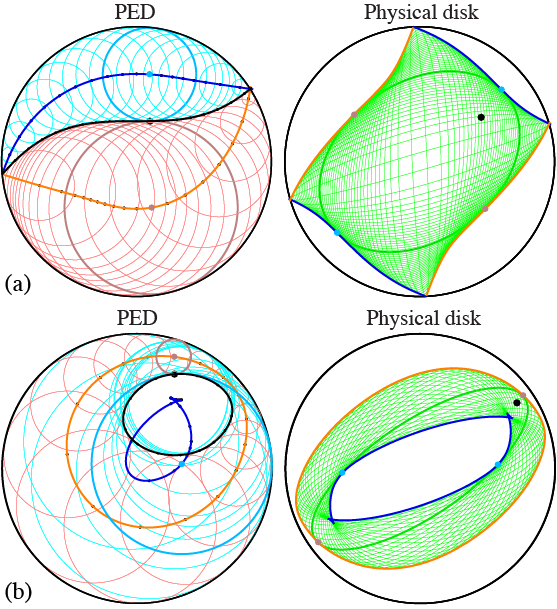}}
\caption{
   Medial axes of Poincar\'e paths in the PED map to caustics in the physical disk.
   Given a Poincar\'e path (thick black curve), one can find two medial axes as the loci of the centers of circles that touch this curve and the unit circle. 
   The mapping $\Z({\bf q})=\pm\sqrt{\Z({\bf t})}$, where ${\bf t}$ are points along the medial axes, corresponds to curves of points ${\bf q}$ that are the caustics of the resulting fields. 
   Note that the caustics in (b) correspond to those in Fig.~\ref{Fig4}(b). 
   } 
\label{Fig6} 
\end{figure}

The geometric relation is easiest to appreciate for a Poincar\'e path in the PED with endpoints at the edge of the disk, such as the one shown in Fig.~\ref{Fig6}(a). 
The corresponding path on the full Poincar\'e sphere is symmetric in the upper and lower hemispheres (projecting to the same curve in the PED). 
The geometric prescription for finding the caustics is then as follows: \\
1) Find the set of circles that are tangent to both the Poincar\'e path and the unit circle. 
There are two such sets on each side of the Poincar\'e path (shown in pale blue and red in Fig.~\ref{Fig6}(a)). 
The centers of each set of circles define a curve equidistant from the Poincar\'e path and the unit circle. 
Each such curve, being equally close to two other curves, is a \emph{medial axis} \cite{Blum} (or {\it topological skeleton}), in terminology borrowed from image analysis. \\
2) Find the caustics by applying the square root map ${\bf z}(\pm\mathcal{Z}^{1/2}(\mb{t}))$ to each medial axis. 
Given that the square root maps each point in the PED onto two points on the physical disk, each medial axis is mapped onto two caustic segments that are identical except for a $180^\circ$ rotation about the origin. 
Therefore, each point along the Poincar\'e path gives rise to two medial axis points and therefore to four caustic points. \\
In cases like that shown in Fig.~\ref{Fig6}(a), where the Poincar\'e path begins and ends at the unit circle, the two medial axes meet at the same endpoints, intersecting each other at right angles. 
Since the square-root mapping is conformal, the caustics in the physical disk also intersect at the disk's edge at right angles.

This construction is also valid for the previous case where a loop is fully within one hemisphere of the Poincar\'e sphere, as shown in Fig.~\ref{Fig6}(b) 
for the same Poincar\'e path as Fig.~\ref{Fig4}. 
Each of the medial axes is now a closed loop, as are the caustics (their square-root images). 
The outer medial axis (orange line), formed by the centers of the (red) circles, is constrained to the annular space between the Poincar\'e path and the unit circle. 
The inner medial axis (blue line) is formed by the centers of the (pale blue) circles that touch the inside of both the unit circle and the Poincar\'e path. 
If, as in this example, there are sufficiently small radii of curvature at some points of the Poincar\'e path, the inner medial axis (and its corresponding caustic) can cross itself and have cusps.

The geometric connection between the path in abstract Poincar\'e space and the beam's caustics in physical space is one of the main results of this work. 
It implies that the caustics of a structured Gaussian beam are composed of two parts that are not mutually independent: 
one can either prescribe a Poincar\'e path and determine the caustics via the medial axes, or instead prescribe one caustic (with the constraint that it must be symmetric under rotations by 180 degrees), then find the corresponding medial axis in the PED, and thus the Poincar\'e path, and then the second medial axis and caustic.

The Poincar\'e path is parametrized as ${\bf s}(\eta)=\left(\cos\theta(\eta)\cos\phi(\eta),\cos\theta(\eta)\sin\phi(\eta)\right)$, so overall the ray family is parametrized by $0 \le \tau,\eta < 2\pi$, topologically corresponding to a torus. 
Families of ellipses in the physical disk such as in Fig.~\ref{Fig4}(b) are projections of this torus, with its outline given by the caustics, consisting of either an outer and an inner loop (as Fig.~\ref{Fig6}(a)), or a quadrangle with corners at the boundary of the physical disk (as Fig.~\ref{Fig6}(b)).
The OPL at $z = 0$ for all the rays, in terms of $\tau$ and $\eta$, may be found in Section S2.

As the Poincar\'e path is a closed loop, wave-optical self-consistency requires that any physical quantity (determined by OPL) must return to its starting point on a circuit of $\eta$. 
This gives a quantization condition around the path, just as our previous condition quantized the orbits.
This condition, whose derivation is in Section S3, is geometrically remarkably simple: the solid angle $\Omega$ on the Poincar\'e sphere enclosed by the Poincar\'e path must be an odd multiple of $2\pi/(N+1)$, namely
\begin{equation}
   \Omega=(2n+1)\frac{2\pi}{N+1}, \qquad n=0,1,...,\lfloor N/2\rfloor,
\label{quant2}
\end{equation}
where $\lfloor N/2\rfloor$ denotes the integer part of $N/2$.

We may appreciate the significance of this by referring to the quantum mechanical picture.
Structured light beams are usually considered as eigenfunctions of some optical operator, such as the OAM operator $\widehat{L} = -\rmi \partial/\partial \phi$ giving the LG modes \cite{Allen}, or the astigmatism operator $\widehat{M} = \tfrac{1}{2}( -\partial^2/\partial x^2 + \partial^2/\partial y^2 + x^2 - y^2 )$ giving the HG modes \cite{OtherPaper}.
In the completely classical, Hamiltonian picture, these quantities are functions of position $\mb{Q}$ and momentum $\mb{P}$, which define families of contours on the Poincar\'e sphere (the sphere of orbits of the isotropic two-dimensional oscillator).
Thus the angular momentum $L$ is simply the height coordinate of the Poincar\'e sphere $\cos\theta$, and for $M$ it is the horizontal coordinate $\sin\theta \cos \phi$ \cite{OtherPaper}; the contours are then circles concentric to the vertical or horizontal axes of the sphere.
The condition (\ref{quant2}) picks out a discrete set of these contours as the Poincar\'e paths, which correspond to the sets of ray families which are consistent with wave optics.
We will discuss the LG and HG modes in much more detail, after we have discussed how to construct approximations to the wave fields from the appropriately quantized ray families.

\section{Ray-based wave field reconstruction}\label{sec:recon}

There are many methods for estimating wave fields based purely on a ray description, which are valid even in the presence of caustics. 
We here use an approach \cite{URBI,URBII,URBIII,SAFEOpEx} in which a complex Gaussian field contribution is assigned to each ray, and the estimate takes the form of a double integral over $\tau$ and $\eta$.
It is shown in Section S5 that the integral in $\tau$ can be evaluated analytically, leading to a field estimate at the waist plane of the form
\begin{eqnarray}
U({\bf x})\!&\approx&\!\frac{kP_0}{\sqrt{2}}\exp\left(\ui\frac{\pi}4\right)\oint A(\eta)\,\sqrt{\cos\theta\frac{\partial\phi}{\partial\eta}+\ui\frac{\partial\theta }{\partial\eta}}\,\,\,U_N\!\left(\frac{\bf x}{Q_0},{\bf v}\right)\nonumber\\
&\times&\!\exp\left(\ui\left\{kL_2-(N+1)\left[T-\frac{\sin(2T)\cos\theta}2\right]\right\}\right) \ud\eta,
\label{SAFEint}
\end{eqnarray}
where ${\bf x}=(x,y)$ is the transverse position at the waist plane, $A(\eta)$ is a non-negative amplitude function weighting the different orbits, ${\bf v}$ is the Jones vector in (\ref{Jones}) parametrized in terms of $\eta$, $T(\eta)$ is given in Section S3, and $U_N$ are Hermite-Gaussian elementary fields evaluated at complex values, defined as
\begin{eqnarray}
U_N(\bar{\bf x},{\bf v}) &=& \frac1{N!}\exp\left(-\frac{N+1}2\right)\,\left(\frac{N+1}2{\bf v}\cdot{\bf v}\right)^{\frac{N}2} \label{SAFEKernel} \\
&\times&\exp\left[-(N+1)|\bar{\bf x}|^2\right]H_N\!\left(\sqrt{\frac{2(N+1)}{{\bf v}\cdot{\bf v}}}\,\bar{\bf x}\cdot{\bf v}\right), \nonumber
\end{eqnarray}
where $H_N$ is the $N^{\rm th}$ order Hermite polynomial and ${\bf v}\cdot{\bf v}=\cos\theta$. 
Up to a complex factor, $U_N$ is the wave contribution corresponding to an elliptical ray orbit specified by the Jones vector ${\bf v}$. 

\begin{figure}[htbp]
\centering
\fbox{\includegraphics[width=\linewidth]{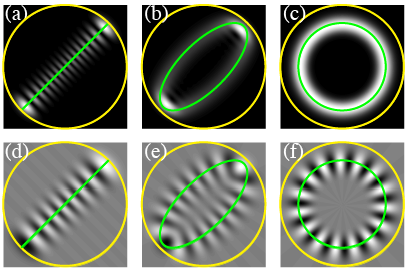}} 
\caption{
   Beam amplitude profiles reconstructed from ray families.
   (a,b,c) Intensities and (d,e,f) real parts of $U_N$, for $N=15$, $\phi=\pi/2$, and (a,d) $\theta=0$, (b,e) $\theta=\pi/4$, and (c,f) $\theta=\pi/2$. 
   The yellow circle indicates the limit of the physical disk, and the ray orbits are shown in green.
} 
\label{SAFEhgs} 
\end{figure} 

Fig.~\ref{SAFEhgs} shows, for several choices of ${\bf v}$, the real part and intensity of each of these orbit contributions, together with the corresponding ray-optical orbit. 
$N$ is evidently the number of phase oscillations around the ellipse. 
In fact, these elementary field contributions are themselves a subset of the HLG beams, which are associated with points over a Poincar\'e sphere \cite{PadgettCourtial,Agarwal,Calvo,HabrakenOL,OtherPaper}. 
However, these contributions are expressed not as a superposition of HG or LG beams but as a single term involving a Hermite polynomial evaluated at a complex argument proportional to the Jones vector. 
The expression in (\ref{SAFEKernel}) provides a general prescription for constructing self-similar beams that are rigorous solutions to the paraxial wave equation, and that have caustics at prescribed locations.

\begin{figure}[htbp]
\centering
\fbox{\includegraphics[width=\linewidth]{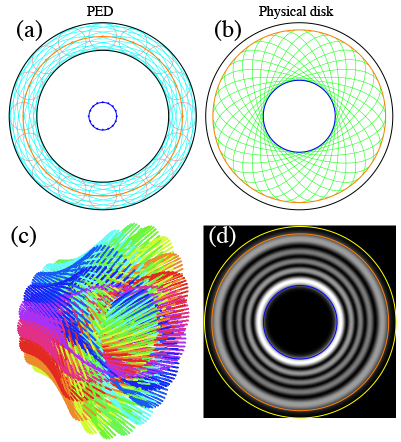}} 
\caption{
   Rays for LG beams with $N=30$ and $n=7$ (so $\ell=22$).
   (a) PED and (b) physical disk.
   In (a), the inner black circle is the Poincar\'e path, and the orange and blue circles are the two medial axes, which map onto the two caustics of the same colors shown in (b) along with some of the elliptical orbits (green). 
   (c) Propagation of the ray family. 
   (d) Wave field intensity with caustics overlaid.} 
\label{LGfigures} 
\end{figure} 

\begin{figure}[htbp]
\centering
\fbox{\includegraphics[width=\linewidth]{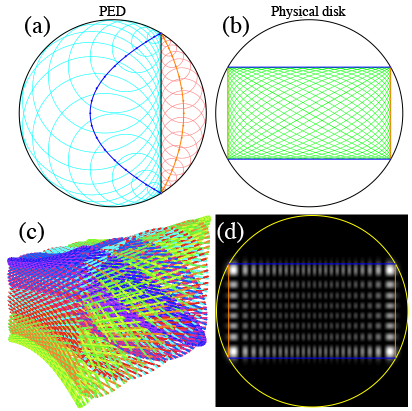}}
\caption{
   Rays for HG beams with $m=23$ and $n=7$ (so $N=30$).
   (a) PED and (b) physical disk.
   In (a), the vertical black line is the Poincar\'e path, and the orange and blue parabolas are its medial axes. 
   These medial axes map onto the straight caustics of the same colors shown in (b) along with some of the elliptical orbits (green). 
   (c) Propagation of the ray family.   
   (d) Wave field intensity with caustics overlaid.} 
\label{HGfigures} 
\end{figure}

\section{Examples: LG, HG and HLG beams} \label{sec:exs}

We now illustrate these ideas for the two most common families of beams of this type, LG and HG beams, as well as for the more general HLG beams. 
LG beams are separable in polar coordinates, and their ray structure was studied by Berry and McDonald \cite{BM}.  
The LG Poincar\'e path in the PED is a circle of radius $r$ centered at the origin (so $\theta=\arccos r$). 
The solid angle enclosed by this circle over the Poincar\'e sphere is $\Omega=2\pi (1-\sin\theta)=2\pi[1-(1-r^2)^{1/2}]$, which is quantized according to (\ref{quant2}), such that $r$ can only take the values
\begin{equation}
   r = \frac{\sqrt{2N+1+4n(N-n)}}{N+1},
   \label{rquantLGHG}
\end{equation}
for $n=0,1,...,\lfloor N/2\rfloor$. 
The medial axes, equidistant from the unit circle and the Poincar\'e path, must also be circles centered at the origin, but with radii $(1\pm r)/2$. 
Following the square root map onto the physical disk, the two caustics are circular as well, with radii $Q_0 \sqrt{(1\pm r)/2}$. 
More details about the ray description of these beams are given in Section S6, where it is also shown that, remarkably, the wave field estimate in (\ref{SAFEint}) actually yields the exact form for LG beams. 
Fig.~\ref{LGfigures} shows the Poincar\'e and physical discs for these beams, including the Poincar\'e path, medial axes, caustics, and elliptical orbits, as well as the ray structure of the beam and the intensity cross section. 
In terms of the operator picture, all physical quantities (Poincar\'e path, medial axes, caustics) must be rotation invariant, and the path quantization gives the usual angular momenta quanta $-N\le \ell \le N$, quantized in integers (in steps of 2).

We now consider HG beams, which are separable in Cartesian coordinates. 
The Poincar\'e path is a straight line terminating at the edge of the PED, as shown in Fig.~\ref{HGfigures}(a). 
Since the PED is a projection of the sphere onto its equatorial plane, the Poincar\'e path on the sphere is a circle centred at the $s_1$ axis with radius $r$, equal to half the length of the straight line, is also quantized according to (\ref{rquantLGHG}) \cite{OtherPaper}. 
By simple geometry, both medial axes are confocal parabolas with foci at the origin, which intersect each other and the Poincar\'e path at the edge of the disk. 
These parabolic medial axes are shown in Fig.~\ref{HGfigures}(a) as blue and orange curves. 
The caustics (square roots of the parabolas) are straight lines, as shown in Fig.~\ref{HGfigures}(b): the first medial axis maps onto two vertical caustic lines (orange) at $x=\pm Q_0\{[1+(1-r^2)^{1/2}]/2\}^{1/2}=\pm Q_0\{(2m+1)/[2(N+1)]\}^{1/2}$ where $m = N-n$, while the second maps onto two horizontal caustic lines (blue) at $y=\pm Q_0\{[1-(1-r^2)^{1/2}]/2\}^{1/2}=\pm Q_0\{(2n+1)/[2(N+1)]\}^{1/2}$. 
Thus the caustics form a rectangle enclosing the rays. 
Further details on the ray parametrization, and a proof that the wave field construction in (\ref{SAFEint}) also gives the exact form for the HG beams, are in Section S7. 
The ray structure and intensity distributions are shown in Figs.~\ref{HGfigures}(c) and (d). 
HG beams are eigenfunctions of the aberration operator $\widehat{M}$ \cite{OtherPaper}, whose eigenvalues $m - n$ are algebraically identical to those of the angular momentum operator.

\begin{figure}[htbp]
\centering
\fbox{\includegraphics[width=\linewidth]{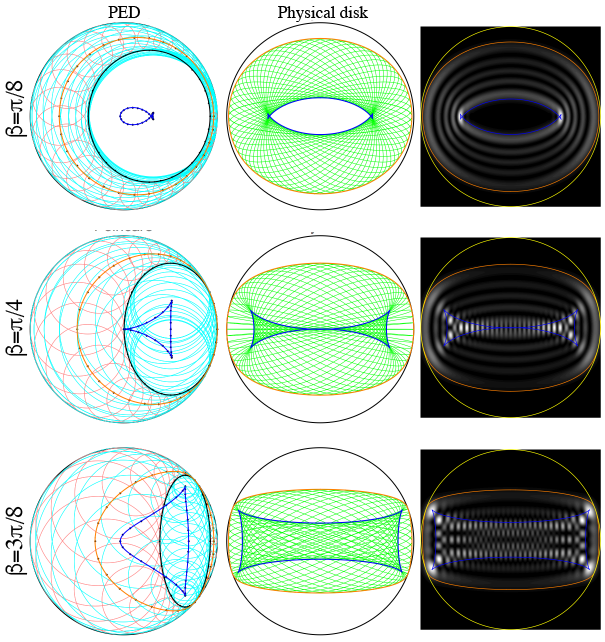}}
\caption{
   PED and physical disk for HLG beams corresponding to $N=30$, $n=4$, and three different angles of rotation $\beta$ in the Poincar\'e sphere. 
   Also shown are the resulting intensity profiles and the ray-optical caustics overlaid.
   The ray families clearly correspond to different projections of a torus, and the brightest parts of the intensities occur in close vicinity of the caustics.
   } 
\label{AVfigs} 
\end{figure}

Finally, we consider HLG beams, which are realized by conversion of HG or LG beams through simple combinations of cylindrical lenses \cite{AV,Beijersbergen} or equivalent SLM implementations, which amount to rotations of the Poincar\'e sphere about an axis in the equatorial plane, but which cannot be expressed simply in any separable coordinate system. 
As for LG and HG beams, for HLG beams the Poincar\'e path on the sphere is a (planar) circle whose radius $r$ is quantized according to (\ref{rquantLGHG}). 
However, the centre of this circle can be at any angle $\beta$ with respect to the vertical $s_3$ axis. 
We assume for simplicity that the centre lies in the $s_1s_3$ plane, so that $\beta=0$ gives LG beams while $\beta=\pi/2$ reduces to HG beams separable in $x$ and $y$. 
Projected onto the equatorial disk, the Poincar\'e path is an ellipse centered at $([1-r^2]^{1/2}\,\sin\beta,0)$ and with minor and major semi-axes given by $r\cos\beta$ and $r$ respectively, as shown in the left column of Fig.~\ref{AVfigs}. 
The medial axes and hence the caustics (shown in the figure's second column) can be found in parametric form, and are not conic sections. 
Similarly, the wave fields are no longer separable in a coordinate system, but they can still be computed from (\ref{SAFEint}). 
Fig.~\ref{AVfigs} illustrates these beams for three values of $\beta$ intermediate between the LG and HG limits. 

In the operator picture, these beams are eigenfunctions of $\widehat{L}\cos\beta + \widehat{M}\sin\beta$, which have the same integer eigenvalues by the hidden symmetry of the isotropic 2-dimensional harmonic oscillator (corresponding, in the ray picture, to rotating the spherical cap in the $s_1s_3$ plane). 
Since the operator is linear in the coordinates of the Poincar\'e sphere's space, the Poincar\'e path is a circle with uniform weight. This simplicity of the HLG family explains why the ray-based field estimate in (\ref{SAFEint}) actually yields the known exact eigenstates of the operators.

\section{Gouy and Pancharatnam-Berry phases}\label{GGP}

In addition to revealing the hidden geometry behind the caustic structure of structured Gaussian beams, the description presented here provides a simple, ray-based explanation for their Gouy and Pancharatnam-Berry phase shifts. 
These two phase shifts turn out to correspond to shifts in each of the two ray parameters, $\tau$ and $\eta$, as follows.

Consider first the case of the Gouy phase shift. 
As shown in (\ref{rayGouy}), propagation in $z$ preserves the ray structure up to a shift $\tau\to\tau+\zeta$, where $\zeta=\arctan(zP_0/Q_0)$. 
Thus, any ray initially at a given location when $z = 0$ is replaced, after propagation, by another one from the same orbit, whose value of $\tau$ is larger by an amount $\zeta$. 
Since a variation in $\tau$ of $2\pi$ corresponds to a path length of $(N+1)\lambda$, this shift in $\tau$ by $\zeta$ amounts to a change in path length of $(\zeta/2\pi)(N+1)\lambda$, and hence to a phase of $(N+1)\arctan(zP_0/Q_0)$, namely the standard Gouy phase for a beam of this type. 
This effect can be appreciated from Fig.~\ref{Fig1}: all rays have roughly the same length. 
However, the ray that touches a given point it the orbit (say, a vertex of the ellipse) at the initial plane is not the same as the one that touches the same point at the final plane. 
The total phase difference is then not only due to the length of the rays but also to the OPL difference between the two rays in question.

\begin{figure}[htbp]
\centering
\fbox{\includegraphics[width=\linewidth]{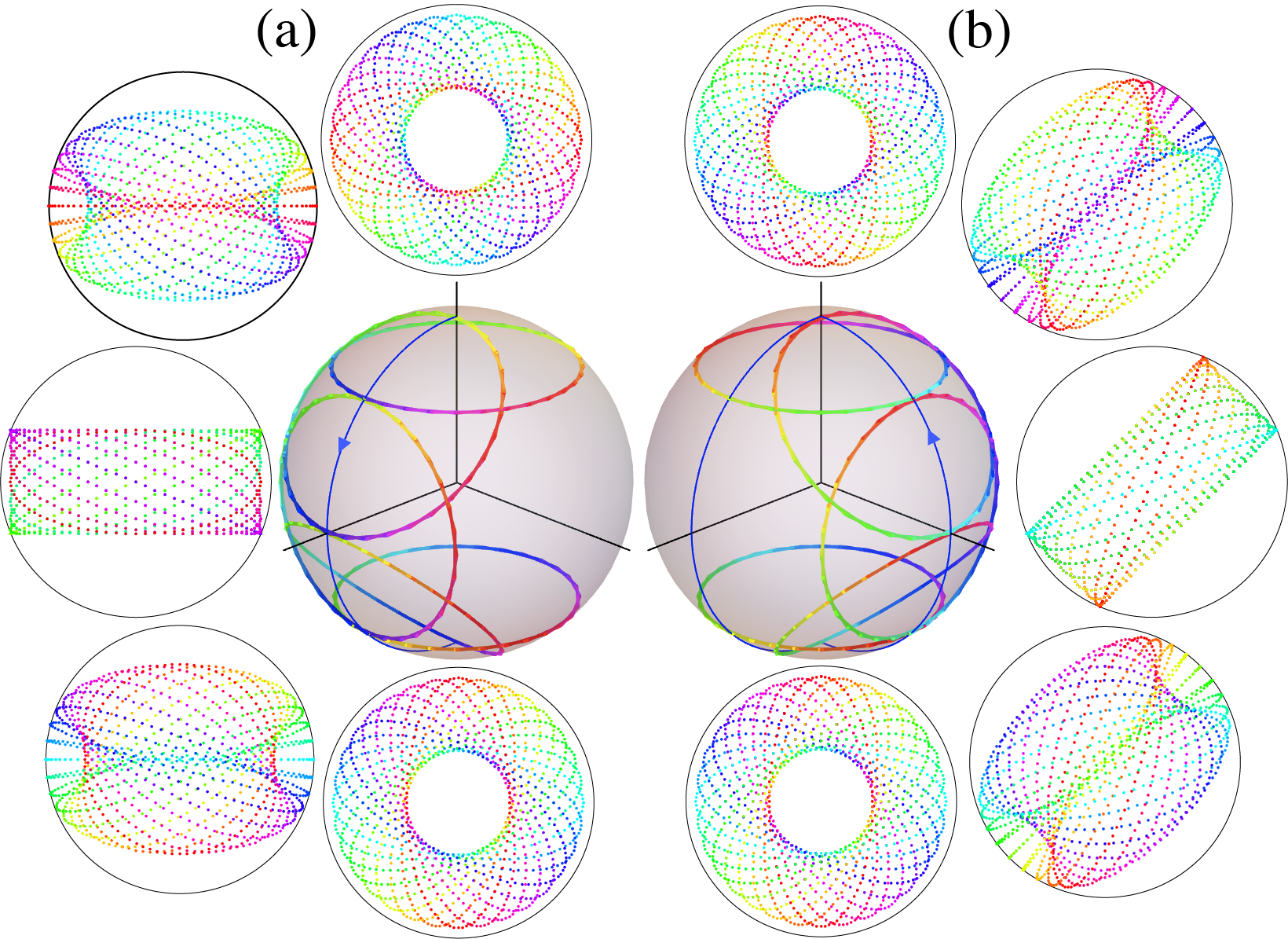}}
\caption{
   Illustration of geometric phase as a cycling of orbits. 
   Part (a) shows the transformation of a LG beam with positive OAM transformed onto one with negative OAM by following a meridional path in the $s_1s_3$ plane over the Poincar\'e sphere. 
   Five stages of this path are shown explicitly, including a HG beam at the equator.
   Part (b) shows the transformation of the beam back onto its initial configuration through a different meridional path. 
   Note that the ray configurations are rotated by $\pi/4$ with respect to those on the left. 
   While the final beam has the same shape as the initial one, the orbits (identified by color) are rotated, resulting on a geometric phase.
   } 
\label{Geom} 
\end{figure}

The geometric phase for beams of the HLG family under astigmatic transformations has been studied in algebraic terms by exploiting the analogy with 2D quantum harmonic oscillators \cite{Calvo,HabrakenSPIE,HabrakenOL,HabrakenJMP} and verified experimentally for low-order beams \cite{GalvezJOSAA,GalvezPRL,MilionePhase}. 
Consider subjecting a HG, LG, or more general HLG beam to a series of optical transformations that rotate the Poincar\'e sphere around an axis within the $s_1s_2$ plane (through a suitable combination of cylindrical lenses) or around the $s_3$ axis (through a beam rotator such as a pair of Dove prisms or periscopes). 
By choosing the sequence of transformations appropriately, the circular Poincar\'e path for the beam can be brought back to its initial position after its center traced a trajectory over the Poincar\'e sphere. 
However, it is easy to see that each point within the Poincar\'e path does not necessarily fall back onto its initial position; rather, the final state of the circle is generally rotated around its axis with respect to the initial one by some angle $\Theta$, depending on the trajectory followed. 
If this trajectory is composed only of segments of great circles, as that shown in Fig.~\ref{Geom}, 
then the angle $\Theta$ equals the solid angle subtended by the path. 
In other words, this transformation reduces to a shift $\eta\to\eta-\Theta$. 
Such rotation results in a phase shift for the beam that can be considered as a geometric phase, because it is not related to a change in the OPL of each ray, but to a cycling of the roles that different rays (and indeed orbits) play within the pattern.  
As stated in Section S3, the phase due to a complete rotation of the Poincar\'e loop is $k\Delta L_2=(N-2n)\pi$, so the corresponding geometric phase is $(N-2n)\Theta/2=\ell\Theta/2$, where $\ell$ is the OAM label of the LG beam within the set.

In summary, the phase accumulated under propagation can be separated into a ``dynamic'' phase, due to the path length traced by each ray, and a Gouy phase, due to the cycling of rays within each orbit. 
If additionally the beam is subjected to a series of transformations that rotate the Poincar\'e sphere but that bring the beam back to its original shape, there is a third, geometric component of the phase, due to the shifting of orbits within the beam structure. 
However, note that while the dynamic and Gouy phases apply to any beam, the general geometric phase can only be achieved for HLG beams, given the rotational symmetry of their Poincar\'e path. 
For beams whose Poincar\'e paths have $M$-fold symmetry around an axis, a more restricted version of the same phenomenon is possible. 

\section{Other separable self-similar beams as limiting cases, and self-healing}\label{sec:sepself}

Despite them not being explicitly Gaussian, other types of propagation-invariant  
beams, such as Bessel \cite{Eberly,Durnin}, Mathieu \cite{Mathieu}, Airy \cite{BerryBalazs,Christodoulides}, and parabolic \cite{parabolic} beams, correspond to limits of the structured Gaussian beams described here. 
These other beams are idealized solutions that involve infinite power, corresponding to the limit $N \to \infty$ in particular regions of the physical disk. 
That is, the ray families are open rather than closed loops.

Bessel and Mathieu beams correspond to a small neighbourhood of the origin of the physical disk, and the outer radius of the PED.
For Bessel beams, the Poincar\'e path is a circle centered at the origin and whose radius is nearly equal (or equal) to unity, so that one medial axis is a small circle (or point) centered at the origin, and so is the inner caustic. 
Mathieu beams use the same construction, except that the large circular Poincar\'e path is shifted slightly from the origin but still fits within the PED.
This shift de-centers the small inner medial axis, and causes the resulting inner caustic to be elliptic. 
(A shift larger than the difference between unity and the path's radius would make the inner caustic hyperbolic.)
If instead the centred Poincar\'e path is slightly deformed into an ellipse, the inner caustic becomes an astroid, as in beams produced by misaligned axicons \cite{SabinoCaustic}.

\begin{figure}[htbp]
\centering
\fbox{\includegraphics[width=\linewidth]{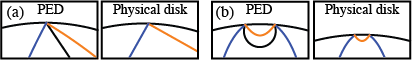}}
\caption{ 
   Relevant segments of the Poincar\'e path and medial axes over a peripheral segment of the PED and the corresponding caustics over a peripheral segment of the physical disk, for (a) a general asymmetric Airy beam and (b) a parabolic beam.} 
\label{Airy} 
\end{figure}

In the case of Airy and parabolic beams, on the other hand, one must focus at a small region at the edge of the PED and physical disk. 
Airy beams are the limit of the intersection of two medial axes (and caustics) when a locally straight Poincar\'e path touches the edge of the PED.
This geometry is shown in Figure \ref{Airy}(a).
The angle of this intersection determines the ratio of the spacing of the intensity lobes along the two caustic sheets. 
Parabolic beams result when the Poincar\'e path is a very small circular segment starting and ending at the edge of the PED, leading to two sets of parabolic caustics, as shown in Fig.~\ref{Airy}(b).

Propagating self-similar beams are often referred to as ``self healing''; if an obstacle blocks a limited part of the beam in one plane, the blocked intensity features reappear as $z$ increases.   
The effect of the block can be described to first order in terms of the ray-optical shadow projected by the obstacle, that is, the suppression of a subset of the rays composing the field.  
Self healing (which can occur more generally \cite{ring:airy,ring:thesis}) is then easily explained in terms of the cycling of rays within each ray orbit under propagation: the blocked rays are replaced by other rays leading to the same local ray structure.
However, it is clear that rather than ``healing'', the beam's ``wound'' is simply transferred to a different part of its transverse profile. 
For beams such as Airy or Bessel beams, the idealized ray family is open, so the shadow is ultimately lost in an infinite reservoir of rays away from the region where the main intensity features are located. 
This is not the case for structured Gaussian beams whose ray family is compact.

Due to the rotational symmetry of their Poincar\'e path, HG, LG, and more general HLG beams can undergo local ``healing'' not only through shifts in $\tau$ under propagation but also through shifts in $\eta$ due to rotations of the Poincar\'e sphere caused by the optical transformations discussed in Section~\ref{GGP}. 
Such shifts would have a similar effect of displacing the blocked regions within the beam's profile. 
By further abusing the already imperfect ``healing'' metaphor, this effect could be called ``assisted healing''.

\section{Concluding remarks}\label{sec:disc}

We proposed a ray-based description of structured Gaussian beams that reveals hidden geometrical restrictions in their spatial structure, particularly their caustics. 
Further, the Gouy and geometric phases that can be accumulated under propagation were also given simple explanations in terms of rays and their quantization. 
The description given here is based on the partition of the two-parameter ray family, one parameter giving rays around orbits with an elliptical cross-section, and the other defining a curve on the Poincar\'e sphere representing the elliptic ray family.
This develops previous work also employing Poincar\'e spheres to characterize the modal structure of HLG beams \cite{PadgettCourtial,Agarwal,GalvezJOSAA,GalvezPRL,Calvo,HabrakenSPIE,HabrakenOL,HabrakenJMP}. 
However, unlike these previous studies, where each beam is associated with a point on the Poincar\'e sphere, in our more general construction the beam is associated with a curve on the Poincar\'e sphere. The shape of this extended curve not only determines the shape of the beam but also explains (and restricts) the geometric phase resulting from beam transformations.

The approach given here also differs from other ray-based studies of structured Gaussian beams. For example, Gaussian beams have been described as bundles of complex rays \cite{Kravtsov,Keller,Deschamps}, as opposed to the real rays used here. Similarly, ray-like descriptions of LG and Bessel beams have been given in terms of Wigner functions \cite{Barnett}, but such a description uses all rays in phase space rather than a two-parameter family, so the concept of caustic is absent and the representation in the Poincar\'e sphere is not compatible with that treatment. Finally, descriptions also exist in terms of curved flux lines rather than rays \cite{BM}.

In the complementary operator picture of our approach \cite{OtherPaper}, there is a spin vector-like operator on the Poincar\'e sphere for which these HLG beams, described by circles whose centers are given by the vector direction of the operator.
The operator approach, built around the $\mathrm{su}(2)$ Poisson algebra of the two-dimensional harmonic oscillator, reveals the algebraic connection between structured Gaussian beams, the classical and quantum harmonic oscillator and the Poincar\'e sphere, contrasting with the semiclassical approach here.

Although our focus here has been the particular examples of HG, LG and HLG beams, the ray-based approach can be applied to any beam with a Gaussian envelope with a well-defined Gouy phase -- in fact, the approach allows such structured Gaussian beams to be designed from almost arbitrary paths on the Poincar\'e sphere satisfying the quantization condition (\ref{quant2}).
One obvious structured Gaussian family we have not explored here is the Ince-Gaussian beams \cite{Ince1,Ince2}, which also interpolate between HG and LG beams but which are separable in elliptic coordinates.
From the other separable beams considered here, one might expect the caustics of Ince-Gaussian ray families to be confocal ellipses and hyperbolas.
Indeed this is the case, and we defer a full discussion to a later article.
Although no other separable Gaussian beam families exist \cite{Boyer}, the freedom of choice of curves on the Poincar\'e sphere allows a huge variety of Gaussian beams with new and unfamiliar properties to be designed.

\section*{Funding Information}
The Leverhulme Trust (MRD), and the National Science Foundation (PHY 1507278) (MAA).

\section*{Acknowledgments}

We thank Michael Berry and John Hannay for useful comments.

\appendix

\renewcommand\figurename{Supplementary Figure}
\setcounter{figure}{0}

\section{Geometric relation between Poincar\'e path and beam caustics}
\label{sec:MedialAxes}
\begin{figure}[htbp]
\centering
\fbox{\includegraphics[width=\linewidth]{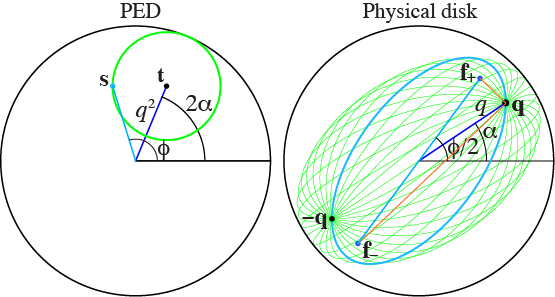}}
\caption{The points in the PED corresponding to all ellipses crossing a prescribed point ${\bf q}$ in the physical disk trace out a circle that touches the edge of the PED and that is centered at the point ${\bf t}$, which is given by the quadratic conformal map of ${\bf q}$ following Eq.~(\ref{Zmaptq}).} 
\label{Fig3} 
\end{figure}

It was shown in Section 4 of the main text 
that a point ${\bf s}$ in the PED corresponds to an ellipse (a curve) in the physical disk through a square-root mapping onto the ellipse's foci and a normalization condition for the ellipse's size. Let us now study the reciprocal map: if we prescribe a point ${\bf q}$ in the physical disk, what is the curve in the PED composed of all points ${\bf s}$ corresponding to ellipses that contain ${\bf q}$? It turns out that this curve is a circle, as we now show. Figure~\ref{Fig3} shows a point ${\bf q}=(q\cos\alpha,q\sin\alpha)$ within the physical disk, as well as several elliptical orbits that contain it. A representative orbit is highlighted in blue. We now use the fact that the sum of the distances (labeled by red lines in Fig.~\ref{Fig3}) from ${\bf q}$ (or any other point along the ellipse) to each of the foci equals the long axis of the ellipse (in this case $2\cos\phi/2$), that is,
\be
|{\bf q}-{\bf f}_+|+|{\bf q}-{\bf f}_-|=2\cos\frac{\phi}2.
\ee
After some manipulation and using the fact that $|{\bf f}_\pm|^2=\cos\theta$, this equation can be reduced to
\be
2q^2\left[1-\cos\theta\cos(\phi-2\alpha)\right]=\sin^2\theta.
\ee
By using ${\bf s}=(\cos\theta\cos\phi,\cos\theta\sin\phi)$, we can write this expression as
\be
\left|{\bf s}-{\bf t}\right|=1-|{\bf t}|.\label{circle1}
\ee
where ${\bf t}=(q^2\cos2\alpha,q^2\sin2\alpha)$. That is, the curve in the PED described by the points ${\bf s}$ corresponding to all the ellipses that include a given point ${\bf q}$ in the physical disk is a circle centered at the point ${\bf t}$, which is related to ${\bf q}$ through a quadratic conformal map of the form
\be
\Z({\bf t})=[\Z({\bf q})]^2.
\label{Zmaptq}
\ee
Further, since the radius of the circle is one minus the distance of its center to the origin, the circle always touches the edge of the PED (a unit disk), as shown in Fig.~\ref{Fig3}.

\begin{figure}[htbp]
\centering
\fbox{\includegraphics[width=\linewidth]{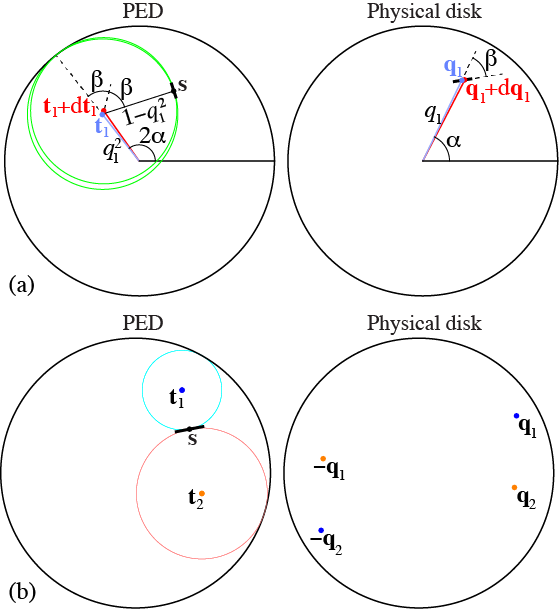}}
\caption{(a) Two endpoints of an infinitesimal caustic segment, ${\bf q}_1$ and ${\bf q}_1+\ud{\bf q}_1$, map onto points ${\bf t}_1$ and ${\bf t}_1+\ud{\bf t}_1$ which are the centers of two circles whose intersection location ${\bf s}$ and direction define an infinitesimal segment of the Poincar\'e path. (b) Given a point ${\bf s}$ and local direction of the Poincar\'e path, there are two circles that are tangent to it and that touch the unit circle. The centers of these circles, ${\bf t}_1$ and ${\bf t}_2$, map onto four caustic points in the Physical disk, $\pm{\bf q}_1$ and $\pm{\bf q}_2$, according to Eq.~(\ref{Zmaptq}).} 
\label{Fig5} 
\end{figure}

The above construction allows answering the question of which infinitesimal Poincar\'e path segment gives rise to a desired caustic segment. Consider an infinitesimal caustic segment joining two points ${\bf q}_1$ and ${\bf q}_1+\ud{\bf q}_1$ over the physical disk, shown in Fig.~\ref{Fig5}(a). These two points correspond to two circles that touch the edge of the PED and that are centered at points ${\bf t}_1$ and ${\bf t}_1+\ud{\bf t}_1$, respectively. Note that, given that the quadratic map from ${\bf q}_1$ to ${\bf t}_1$ in Eq.~(\ref{Zmaptq}) is conformal except at the origin, the angle $\beta$ between ${\bf q}_1$ and $\ud{\bf q}_1$ is the same as that between ${\bf t}_1$ and $\ud{\bf t}_1$. The circles centered at ${\bf t}_1$ and ${\bf t}_1+\ud{\bf t}_1$, respectively, intersect at two points, one of which is always at the edge of the disk. The second intersection, labeled as ${\bf s}$ in Fig.~\ref{Fig5}(a), is at the mirror image of the first intersection with respect to the line containing ${\bf t}_1$ and $\ud{\bf t}_1$. Therefore, this second intersection is located at ${\bf s}={\bf t}_1+(1-q_1^2)[\cos2(\alpha-\beta),\sin2(\alpha-\beta)]$. By construction, ${\bf s}$ is then the point in the PED associated with an ellipse in the physical disk that contains the desired caustic segment, {\it i.e.}, it is a point along the Poincar\'e path.  
However, it is not sufficient to prescribe ${\bf s}$; the local direction of the Poincar\'e path at ${\bf s}$ must also be specified, so that the ellipses corresponding to neighboring point are nearly tangent to the caustic segment too. This local direction is simply the direction of the circle segments at the intersection, which from Fig.~\ref{Fig5}(a) is seen to correspond to the angle $2(\alpha-\beta)+\pi/2$ with respect to the $s_1$ (horizontal) axis.

Consider now the case where an infinitesimal segment of the Poincar\'e path at a point ${\bf s}$ is prescribed. Figure~\ref{Fig5}(b) shows that there are two circles that touch tangentially both this segment and the unit circle, whose centers are labelled as ${\bf t}_1$ and ${\bf t}_2$. Each of these points is related to a different caustic, and maps (under the square root map) to two symmetrically distributed points along it, $\pm{\bf q}_1$ and $\pm{\bf q}_2$. As we move along the Poincar\'e path, the points ${\bf t}_1$ and ${\bf t}_2$ (which are the circles constrained to touch the unit circle and the Poincar\'e path) trace what is known as a medial axis. The caustics result from applying the square root mapping to these two medial axes.

\section{OPL for the complete ray family}
\label{OPLall}
We now derive the parametrized OPL for the complete ray family. Let the Poincar\'e path be parametrized as ${\bf s}(\eta)=[\cos\theta(\eta)\cos\phi(\eta),\cos\theta(\eta)\sin\phi(\eta)]$, and let $L_2(\eta)$ be the OPL of a subset of rays along some curve in the physical disk that touches all the orbits. The increment in OPL for the rays along this curve is again given by $\ud L_2={\bf P}\cdot\ud{\bf Q}$. By using ${\bf Q}=Q_0{\bf q}$ and ${\bf P}=P_0\,\partial{\bf q}/\partial\tau$, this increment can be written as
\bea
\ud L_2={\bf P}\cdot\frac{\partial{\bf Q}}{\partial\eta}\ud\eta=Q_0P_0\frac{\partial{\bf q}}{\partial\tau}\cdot\frac{\partial{\bf q}}{\partial\eta}\ud\eta.\label{2ndOPL}
\eea
It is convenient to integrate this equation along one of the caustics, so that $\partial{\bf q}/\partial\eta$ is parallel to $\partial{\bf q}/\partial\tau$ and their dot product equals the product of the magnitudes. Also, from Eqs.~(1) and (3) we find the relation $|{\bf P}/P_0|^2+|{\bf Q}/Q_0|^2=1$, from which we get $|\partial{\bf q}/\partial\tau|=\sqrt{1-|{\bf q}|^2}$. With this, Eq.~(\ref{2ndOPL}) can be simplified to
\be
\ud L_2=Q_0P_0\sqrt{1-|{\bf q}_i|^2}\left|\frac{\partial{\bf q}_i}{\partial\eta}\right|\,\ud\eta,\label{2ndOPL2}
\ee
for $i=1,2$. This expression can be mapped onto the PED since points ${\bf q}_i$ along a caustic map onto points ${\bf t}_i$ along a medial axis. 
Since this mapping is quadratic and conformal, it is easy to see that $|\partial{\bf q}_i/\partial\eta|=|\partial{\bf t}_i/\partial\eta|/2\sqrt{|{\bf t}_i|}$, yielding
\be
\ud L_2=\frac{Q_0P_0}2\sqrt{\frac{1-|{\bf t}_i|}{|{\bf t}_i|}}\left|\frac{\partial{\bf t}_i}{\partial\eta}\right|\,\ud\eta,\label{2ndOPL3}
\ee
so that $L_2$ can be written as an integral along a medial axis as
\be
L_2(\eta) =\frac{Q_0P_0}2\int
^{\eta}\sqrt{\frac{1-|{\bf t}_i(\eta')|}{|{\bf t}_i(\eta')|}}\,\left|\frac{\partial{\bf t}_i}{\partial\eta'}(\eta')\right|\,\ud\eta'.\label{2ndOPL4}
\ee

The function that describes the OPL for any ray in the family is then equal to the OPL up to the orbit containing the ray plus the OPL between the ray in question and the ray in the orbit touching the caustic:
\bea
\!\!\!\!L(\tau,\eta)\!\!&=&\!\!L_2(\eta)+L_1(\tau)-L_1[T(\eta)] \label{OPLforallrays}\\
&+&\!\!L_2(\eta)\!+\!\frac{Q_0P_0}2\!\left[\tau-T+\frac{\sin(2T)-\sin(2\tau)}2\!\cos\theta\right]\!, \nonumber
\eea
where $T(\eta)$ is the value of $\tau$ for the orbit labeled by $\eta$ corresponding to the ray that touches the reference caustic. This value can be found as
\bea
T(\eta)&=&\arctan\left\{\left[\begin{array}{cc}\sec(\theta/2)&0\\0&\csc(\theta/2)\end{array}\right]\cdot\mathbb{R}\left(\frac{\phi}2\right)\cdot{\bf q}_i(\eta)\right\}\nonumber\\
&=&{\rm arg}\left[{\bf v}(\eta)\cdot\left(\begin{array}{cc}0&\ui \\-\ui &0\end{array}\right)\cdot{\bf q}_i(\eta)\right],
\eea
where $\mathbb{R}(\phi/2)$ is a rotation matrix by an angle $\phi/2$, ${\bf q}_i(\eta)$ is the parametrized caustic, $\arctan({\bf u})=\arctan(u_x,u_y)$ is the two-parameter arctangent function, and $\theta(\eta)$ and $\phi(\eta)$ follow from the specification of the Poincar\'e path ${\bf s}(\eta)=[\cos\theta(\eta)\cos\phi(\eta),\cos\theta(\eta)\sin\phi(\eta)]$.

\section{Quantization of the enclosed solid angle}
\label{sec:secondquantization}
We now derive the second quantization condition for the ray family. For simplicity, consider a case where the Poincar\'e path traces a full loop within the PED, as in Fig.~4(b) of the main text. 
Let $\Delta L_2$ be the total increment of optical path that results from tracing a complete medial axis ({\it e.g.}, the thick orange loop inside the PED in Fig.~4(b)):
\bea
\Delta L_2&=&L_2(2\pi)-L_2(0)\nonumber\\
&=&\frac{Q_0P_0}4\left(2\int_{0}^{2\pi}\sqrt{\frac{1-|{\bf t}|}{|{\bf t}|}}\,\left|\frac{\partial{\bf t}}{\partial\eta}\right|\,\ud\eta\right).\label{2ndOPL5}
\eea
It is shown in the next Section of this document that the quantity in parentheses equals the solid angle $\overline{\Omega}$ enclosed between the Poincar\'e path and the equator over the surface of the Poincar\'e sphere. Therefore, by also recalling the first quantization condition in Eq.~(6), Eq.~(\ref{2ndOPL5}) can be written as 
\be
\Delta L_2=\frac{(N+1)\lambda}{4\pi}\overline{\Omega}=\frac{(N+1)\lambda}{4\pi}(2\pi-\Omega),\label{2ndOPL6}
\ee
where $\Omega$ is the solid angle enclosed by the Poincar\'e path. 

Given the square-root mapping, spanning the complete medial axis amounts to spanning only half of the corresponding caustic (say, the bottom-right half between the two marked dots of the orange caustic loop in the physical disk in Fig.~4(b)). 
That is, $\Delta L_2$ is the optical path difference between two points along the caustic that are symmetrically located around the origin, and that therefore belong to the same orbit (indicated by the thicker green ellipse in the figure). 
Therefore, the phase difference due to $\Delta L_2$ must be consistent with that calculated from the optical path difference of the corresponding endpoint rays along the orbit they belong to ({\it i.e.}, along one half of the thick green ellipse joining the two orange points in Fig.~4(b)), given from Eqs.~(5) and (6) by $L_1(\tau_0+\pi)-L_1(\tau_0)=(N+1)\lambda/2$. Note, however, that when calculating the phase difference from the orbit, one must add by hand an extra phase term resulting from the fact that the ray family touches caustics between the two points. Phases resulting from caustics are known as Maslov phases, and have a magnitude of $\pi/2$. In this case, the total phase due to caustics is $\pi$, since the elliptical segment between the two rays in question touches two caustics (``half'' at each contact point indicated by an orange dot with the caustic in question, and one in-between with the other caustic, indicated by a pale blue dot). The phases along the caustic and the orbit are then consistent if
\be
\frac{2\pi}{\lambda}\Delta L_2+2\pi m=\pi(N+1)+\pi,
\ee
where $m$ is an integer. In other words, after using Eq.~(6) and Eq.~(\ref{2ndOPL5}) of this document, the enclosed solid angle $\Omega$ must take values:
\be
\Omega=(2n+1)\frac{2\pi}{N+1},\,\,\,\,\,n=0,1,...,\lfloor N/2\rfloor,
\label{quant2ap}
\ee
where $n=m-1$. That is, the total solid angle enclosed by the Poincar\'e path must be an odd multiple of $2\pi/(N+1)$.

Let us now show that this quantization for the enclosed solid angle also holds for cases where the path in the PED starts and finishes at the disk's edge, as in Fig.~4(a) Let the starting and finishing points correspond to values $\eta_1$ and $\eta_2$ of the parameter $\eta$. For simplicity, assume that this open path in the PED corresponds to the projection onto the equatorial plane of a closed loop over the surface of the Poincar\'e sphere that occupies both hemispheres and has mirror symmetry with respect to the equatorial plane. In this case, one must verify that the phase due to the optical path difference between the endpoints of the path segment corresponding to the $s_3\ge0$ hemisphere is consistent with that for the path segment corresponding to the $s_3\le0$ hemisphere. Since the orbital angular momentum is reversed between these hemispheres, so is the accumulation of optical path difference. This means that $L(\eta_2)-L(\eta_1)$ for the lower hemisphere has equal magnitude but opposite sign to the corresponding path difference for the upper hemisphere, so that the total optical path accumulated around the loop is $\Delta L_2=2L(\eta_2)-2L(\eta_1)$. Since the closed loop (corresponding to one of the caustics) touches the other caustic twice (once at each equatorial point), an extra Maslov phase of $\pi$ must be included. The quantization condition in this case can then be written as
\be
\frac{2\pi}{\lambda}\Delta L_2+\pi=2\pi m',
\ee
where $m'$ is an integer. However, $L_2(\eta_2)-L_2(\eta_1)$ corresponds to $Q_0P_0/4$ times the solid angle within a hemisphere to one side of the loop (depending on which caustic was used), and therefore $\Delta L_2=2L_2(\eta_2)-2L_2(\eta_1)=Q_0P_0/4(4\pi-\Omega)$, where $\Omega$ is the solid angle enclosed/excluded by the symmetric loop. By now defining $n=N-m'+1$, we find that the quantization condition is indeed the one in Eq.~(8). 
It is easy to see that the same condition would hold for a loop that occupies both hemispheres but is not symmetric. 

Finally, notice that whether the loop is restricted to a hemisphere or not, we have the relation $k\,\Delta L_2=(N-2n)\pi$.

\section{Solid angle interpretation of the integral}
\label{sec:spherearea}
The equation for the circles centered at ${\bf t}$ that touch both the Poincar\'e path and the edge of the PED (the unit circle) is given in Eq.~(\ref{circle1}). This equation can be parametrized as
\be
{\bf s}({\bf t},\xi)={\bf t}+(1-|{\bf t}|){\bf u}(\xi),
\ee
where ${\bf u}(\xi)$ is a unit vector at an angle $\xi$ measured, for convenience, from the direction of ${\bf t}$. As ${\bf t}$ runs along one of the medial axes, the circles advance touching both the unit circle and the Poincar\'e path, as shown in Fig.~4.  

\begin{figure}[htbp]
\centering
\fbox{\includegraphics[width=\linewidth]{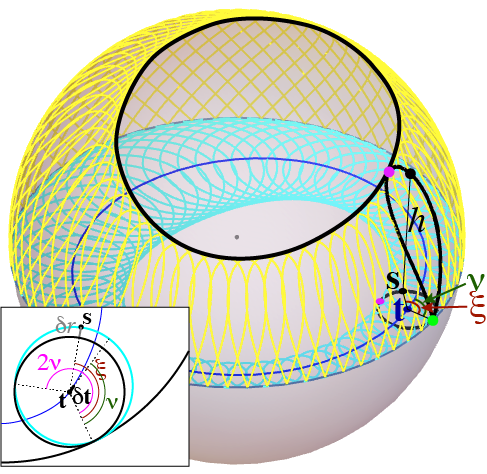}}
\caption{Vertical projections (yellow curves) onto the upper hemisphere of the Poincar\'e sphere of the circles (pale blue) that touch the edge of the equatorial disk and that are centered at the medial axis (blue curve). The inset illustrates the geometry over a segment of the PED.} 
\label{SA} 
\end{figure} 
The projections of these circles over the surface of the Poincar\'e sphere are three-dimensional curves that touch both the Poincar\'e curve and the equator. These curves are shown in yellow in Fig.~\ref{SA}, with one of them highlighted in black. Their projections onto the equatorial plane are the pale blue circles (one also highlighted in black), whose centers ${\bf t}$ trace the medial axis (blue curve).  The solid angle enclosed between the Poincar\'e curve (thick black) and the equator can then be calculated as the solid angle subtended by either the leading or the trailing segments of the projections of the circles joining these curves as ${\bf t}$ runs through the medial axis. While using either the leading or trailing segments alone would give the same result for the integral over the whole medial axis, it would give different forms for the integrands. Using the average of the two options turns out to give a simpler expression, and more importantly, the one we set out to find.

We now calculate the average of the two solid angle elements corresponding to the changes in both the leading and trailing segments of the circle as ${\bf t}$ varies by an infinitesimal amount $\delta{\bf t}$, at an angle $\nu$ with respect to ${\bf t}$. The first step is to calculate the radial displacement of the points in the circle with respect to its center, resulting from this infinitesimal change. This radial displacement is given by
\bea
\delta r(\xi)&=&[{\bf s}({\bf t},\xi)-{\bf s}({\bf t}+\delta{\bf t},\xi)]\cdot{\bf u}(\xi)=\delta{\bf t}\cdot{\bf u}(\xi)-|\delta{\bf t}|\cos\nu\nonumber\\
&=&|\delta{\bf t}|\left[\cos(\xi-\nu)-\cos\nu\right]\nonumber\\
&=&-2|\delta{\bf t}|\sin\left(\frac{\xi}2-\nu\right)\sin\frac{\xi}2.
\eea
As could be expected, this radial displacement vanishes for $\xi$ equal zero and $2\nu$, which correspond to the intersections of the circles at the unit circle and the Poincar\'e curve, respectively, and which are shown as green and purple dots in the highlighted curve in Fig.~\ref{SA}. These two values are the boundaries between the leading and trailing segments of the circle. If we were to integrate this displacement in $\xi$ times the radius $(1-|{\bf t}|)$ over an interval of size $2\pi$,, we would obtain the area of the two complementary crescents enclosed by the initial and final circles. However, what we want to find instead is the solid angle element projected over the surface by this area, which is given by the integral of the magnitude of this radial thickness times the length element $(1-|{\bf t}|)\ud\xi$, divided by an obliquity factor $h(\xi)$:
\be
\delta\overline{\Omega}=\frac12\int_{\xi_0}^{\xi_0+2\pi}\frac{|\delta r(\xi)|(1-|{\bf t}|)}{h(\xi)}\,\ud\xi,
\ee
where $\xi_0$ is an arbitrary limit of integration. Note that we use the absolute value of the radial displacement so that we add rather than subtract the solid angles spanned by the leading and trailing segments of the circles, and that a factor of 1/2 is included for performing an average.  Given the spherical geometry, the obliquity factor $h$ equals the height of the projection of the point in question onto the surface of the sphere:
\bea
h(\xi)&=&\sqrt{1-|{\bf s}(\xi)|^2}\nonumber\\
&=&\sqrt{2|{\bf t}|(1-|{\bf t}|)(1-\cos\xi)}\nonumber\\
&=&2\sqrt{|{\bf t}|(1-|{\bf t}|)}\left|\sin\frac{\xi}2\right|.
\eea
Substituting into the integral for the solid angle element, we get
\bea
\delta\overline{\Omega}\!&=&\!\frac12\sqrt{\frac{1-|{\bf t}|}{|{\bf t}|}}|\delta{\bf t}|\int_{\xi_0}^{\xi_0+2\pi}\!\left|\sin\left(\frac{\xi}2-\nu\right)\right|\ud\xi\!\nonumber\\
&=&\!2\sqrt{\frac{1-|{\bf t}|}{|{\bf t}|}}|\delta{\bf t}|,
\eea
where the result in the integral in the last step is easiest to find by setting $\xi_0=2\nu$. It is remarkable that this result is independent of $\nu$, the local direction of the medial axis with respect to the radial direction. The main result of this Section is then the fact that the integral of the previous result over the complete medial axis equals the solid angle enclosed between the Poincar\'e curve and the equator:
\bea
\overline{\Omega}&=&2\int\sqrt{\frac{1-|{\bf t}|}{|{\bf t}|}}|\ud{\bf t}|
=2\int\sqrt{\frac{1-|{\bf t}|}{|{\bf t}|}}\left|\frac{\partial{\bf t}}{\partial\eta}\right|\ud\eta,
\eea
where $\eta$ is a parameterization of ${\bf t}$ and the integral must extend over the complete medial axis. 

\section{Ray-based field estimates}
\label{sec:SAFEest}
The ray-based wave field estimate in Refs. [48-51] from the main manuscript is constructed as a sum of complex Gaussian field contributions, each centered at a ray, according to
\bea
U({\bf x}) &\approx& \frac{k}{2\pi}\int\int_0^{2\pi}A(\eta)\,\sqrt{\frac{\partial(\Gamma{\bf Q}+\ui {\bf P})}{\partial(\tau,\eta)}}\label{SAFE}\\
& &\times\exp\!\left\{-\frac{k\Gamma}2|{\bf x}-{\bf Q}|^2+\ui k\left[L+({\bf x}-{\bf Q})\cdot{\bf P}\right]\right\}\!\ud\eta\ud\tau,\nonumber
\eea
where ${\bf x}=(x,y)$ are the transverse coordinates at the waist plane, $\Gamma$ is a constant with units of inverse length that regulates the widths of the Gaussian contributions, and $A(\eta)$ (chosen here to be independent of $\tau$) is a non-negative amplitude function indicating the relative importance of the different orbits. The field estimate in Eq.~(\ref{SAFE}) is asymptotically insensitive (for large $k$) to the choice of the scale parameter $\Gamma$. However, for our current purposes, it is convenient to use $\Gamma=P_0/Q_0$ (so that the Gaussian elements have the same width as the elementary Gaussian beam at the waist). 

The Jacobian in Eq.~(\ref{SAFE}) can then be found from Eqs.~(2) and (3) 
to be
\be
\frac{\partial(\Gamma{\bf Q}+\ui {\bf P})}{\partial(\tau,\eta)}=
\ui P_0^2\exp(-2\ui\tau)\Delta(\eta),
\ee
where
\bea
\Delta(\eta)&=&v_x(\theta,\phi)\frac{\partial v_y(\theta,\phi)}{\partial\eta}-v_y(\theta,\phi)\frac{\partial v_x(\theta,\phi)}{\partial\eta}\nonumber\\
&=&\frac12\left(\cos\theta\frac{\partial\phi}{\partial\eta}+\ui\frac{\partial\theta }{\partial\eta}\right),
\eea
with $v_x$ and $v_y$ being the components of the Jones vector in Eq.~(1).  
Equation~(\ref{SAFE}) 
can then be written as
\bea
U({\bf x}) &\approx& \frac{kP_0}{2\pi}\exp\left[\ui\frac{\pi}4-(N+1)\left(\frac{|{\bf x}|^2}{Q_0^2}+\frac12\right)\right]\\
& &\times \int A(\eta)\,\sqrt{\Delta(\eta)}\,{\cal I}({\bf x},\eta) \nonumber \\
& & \times 
\exp\!\!\left(\!\ui\left\{\!kL_2(\eta)\!-\!(N+1)\!\!\left[\!T\!-\!\frac{\sin(2T)\cos\theta}2\!\right]\!\right\}\!\right)\!\ud\eta. \nonumber
\eea
Here, ${\cal I}$ is the integral over the $\tau$ dependent parts, which can be solved analytically:
\bea
{\cal I}({\bf x},\eta)
&=&\int_0^{2\pi}\exp(\ui N\tau)\exp\Bigg[2(N+1)\frac{{\bf x}\cdot{\bf v}}{Q_0}\exp(-\ui\tau)\nonumber\\
& &-\frac{N+1}2\cos\theta\exp(-2\ui\tau)\Bigg]\,\ud\tau\nonumber\\
&=&-\ui\oint_{\rm u.c.}\frac1{u^{N+1}}\nonumber\\
& &\times\exp\left(-\frac{N+1}2\cos\theta\,u^2+2(N+1)\frac{{\bf x}\cdot{\bf v}}{Q_0}\,u\right)\,\ud u\nonumber\\
&=&-\ui\exp\left(\frac{N+1}2\cos\theta\,u_0^2\right)\nonumber\\
& & \times\oint_{\rm u.c.}\frac1{u^{N+1}}\exp\left[-\frac{N+1}2\cos\theta\,(u-u_0)^2\right]\,\ud u\nonumber\\
&=&2\pi\frac{(-1)^N}{N!}\exp\left(\frac{N+1}2\cos\theta\,u_0^2\right)\nonumber\\
& &\times\frac{\ud^N}{\ud u_0^N}\exp\left(-\frac{N+1}2\cos\theta\,u_0^2\right)\nonumber\\
&=&\!\frac{2\pi}{N!}\left(\!\frac{N+1}2\!\cos\theta\!\right)^{N/2}\!\!\!\!\!\!H_N\!\!\left(\sqrt{\frac{2(N+1)}{\cos\theta}}\,\frac{{\bf x}\cdot{\bf v}}{Q_0}\right),
\eea
where in the second step we used the change of variables $u=\exp(-\ui\tau)$ so that the integral is over the unit circle (u.c.) of the complex plane, in the third step we used the shorthand $u_0=2{\bf x}\cdot{\bf v}/[Q_0\cos\theta]$, in the fourth step we applied residue theory, and in the last step we recognized the generating function of the $N$th order Hermite polynomial $H_N$. With this, the field estimate can be written in the form in Eq.~(9).

\section{Laguerre-Gaussian beams}
\label{sec:AppLG}
For LG beams, the Poincar\'e path is a circle of radius $r=\cos\theta$ centered at the origin. For convenience, we choose $\phi(\eta)=2\eta$ (for $\eta$ varying over a range of $\pi$), so that ${\bf s}=(r\cos2\eta,r\sin2\eta)$. The Jones vector then reduces to
\be
{\bf v}=\sqrt{\frac{1+r}2}\left(\cos\eta,\sin\eta\right)+\ui\sqrt{\frac{1-r}2}\left(-\sin\eta,\cos\eta\right),
\ee
so that $v_x\partial v_y/\partial\eta-v_y\partial v_x/\partial\eta=r$. 
The position and momentum vectors are
\bea
{\bf Q}&=&Q_0\sqrt{\frac{1+r}2}\left(\cos\eta,\sin\eta\right)\cos\tau\nonumber\\
&+&Q_0\sqrt{\frac{1-r}2}\left(-\sin\eta,\cos\eta\right)\sin\tau,\label{qLG}\\
{\bf P}&=&-P_0\sqrt{\frac{1+r}2}\left(\cos\eta,\sin\eta\right)\sin\tau\nonumber\\
&+&P_0\sqrt{\frac{1-r}2}\left(-\sin\eta,\cos\eta\right)\cos\tau,\label{pLG}
\eea
Also, it is easy to see that $T(\eta)=0$ since the ellipses touch the outer caustic at their vertices, so the total OPL is given by
\bea
kL(\tau,\eta)&=&k[L_2(\eta)+L_1(\tau)]\nonumber\\
&=&(N+1)\left(\eta\sqrt{1-r^2}+\tau-\frac{\sin2\tau}2r\right),\label{OPLG}
\eea
with the quantization condition in Eq.~(11). 

Let us write ${\bf q}={\bf Q}/Q_0$ in polar coordinates $(q,\varphi)$, which from Eq.~(\ref{qLG}) are given by
\bea
q&=&\sqrt{{\bf q}\cdot{\bf q}}=\sqrt{\frac{1+r\cos2\tau}2}=\sqrt{\frac{1+r-2r\sin^2\tau}2},\\
\varphi&=&\arctan\frac{q_y}{q_x}=\eta+\arctan\left(\sqrt{\frac{1-r}{1+r}}\tan\tau\right).
\eea
The ray parameters can then be expressed in terms of these coordinates as
\bea
\tau&=&\arcsin\left(\sqrt{\frac{1+r-2q^2}{2q^2-1+r}}\right),\\
\eta&=&\varphi-\arctan\left[\sqrt{\frac{(1-r)(1+r-2q^2)}{(1+r)(2q^2-1+r)}}\right].
\eea
From these expressions we see that the OPL in Eq.~(\ref{OPLG}) can be written as a sum of a part dependent purely on $q$ and a part dependent purely on $\varphi$, pointing at the separability of the problem in polar coordinates.

We now evaluate the field estimate in Eq.~(9)  
by writing the normalized position vector ${\bf x}$ also in polar coordinates as ${\bf x}=(Q_0 \bar{\rho},\vartheta)$. By setting $A(\eta)=1$ and defining $\ell=N-2n$,  
Eq.~(9) simplifies to
\bea
U({\bf x})\!\!&\approx&\!\!
kP_0\exp\left(\ui\frac{\pi}4\right)\exp\left(-\frac{N+1}2\right)\frac{\left[(N+1)^2-\ell^2\right]^{\frac{N+1}4}}{2^{\frac N2}N!\sqrt{N+1}}\nonumber\\
& &\times \exp\left[-(N+1)\bar{\rho}^2\right]\exp(\ui \ell\vartheta)\\
& & \times \!\!\!\int_0^{2\pi}\!\!\!\!\!H_{\ell+2n}\!\left[a_-\exp(\ui\eta')\!+\!a_+\exp(-\ui\eta')\!\right]\!\exp\!(\ui \ell\eta')\ud\eta'\!, \nonumber
\eea
where $\eta'=\eta-\vartheta$ and $a_\pm=\sqrt{(N+1)/r}(\sqrt{1+r}\pm\sqrt{1-r})\bar{\rho}/2$. Notice that the range of integration was extended from $[0,\pi]$ to $[0,2\pi]$ by inserting a factor of $1/2$ and using the fact that $\ell$ has the same parity as $N$ and hence the integrand is a combination of even powers of $\exp(\ui\eta')$. This integral can then be solved by using the expansion for Hermite polynomials:  
\bea
&&\int_0^{2\pi} \sum_{i=0}^{n+\lfloor \ell/2\rfloor}\frac{(-1)^i n!2^{\ell+2(n-i)}}{i![\ell+2(n-i)]!}\nonumber\\
&\times&\left[a_-\exp(\ui\eta')+a_+\exp(-\ui\eta')\right]^{\ell+2(n-i)}\exp(\ui \ell\eta')\,\ud\eta'\nonumber\\
&=&\sum_{i=0}^{n+\lfloor \ell/2\rfloor}\frac{(-1)^i n!2^{\ell+2(n-i)}}{i![\ell+2(n-i)]!}\sum_{n'=0}^{\ell+2(n-i)}\left(\begin{array}{cc}\ell+2(n-i)\\n'\end{array}\right)\nonumber\\
&\times&a_-^{\ell+2(n-i)-n'}a_+^{n'}\int_0^{2\pi}\exp[2\ui(\ell+n-i-n')\eta']\,\ud\eta'\nonumber\\
&=&2\pi\sum_{i=0}^{n+\lfloor \ell/2\rfloor}\frac{(-1)^i n!2^{\ell+2(n-i)}}{i![\ell+2(n-i)]!}\left(\begin{array}{cc}\ell+2(n-i)\\ \ell+n-i\end{array}\right)a_-^{n-i}a_+^{\ell+n-i}\nonumber\\
&=&2\pi\sum_{i=0}^{n}\frac{(-1)^i n!2^{\ell+2(n-i)}}{i!(\ell+n-i)!(n-i)!}a_-^{n-i}a_+^{\ell+n-i}\nonumber\\
&=&2\pi(-1)^{n}\left(\sqrt{1+r}+\sqrt{1-r}\right)^{\ell}\left(\sqrt{\frac{N+1}r}\bar{\rho}\right)^{\ell}\nonumber\\
&\times&\sum_{i'=0}^{n}\frac{(-1)^{i'}n!}{i'!(n-i')!(\ell+i')!}\left[2(N+1)\bar{\rho}^2\right]^{i'}\nonumber\\
&=&2\pi\frac{(-1)^{n}n!}{(\ell+n)!}\left(\sqrt{1+r}+\sqrt{1-r}\right)^{\ell}\left(\sqrt{\frac{N+1}r}\bar{\rho}\right)^{\ell}\nonumber\\
&\times&\sum_{i'=0}^{n}\frac{(-1)^{i'}}{i'!}\left(\begin{array}{cc}n+\ell\\n-i'\end{array}\right)\left[2(N+1)\bar{\rho}^2\right]^{i'}
\nonumber\\
&=&2\pi\frac{(-1)^{n}n!}{(\ell+n)!}\left(\sqrt{1+r}+\sqrt{1-r}\right)^{\ell} \left(\sqrt{\frac{N+1}r}\bar{\rho}\right)^{\ell},\nonumber\\
&\times&L_n^{(\ell)}\left[2(N+1)\bar{\rho}^2\right],
\eea
where we used the substitution $i'=n-i$ and in the last step we recognized the series definition of the associated Legendre polynomial $L_n^{(\ell)}$. The substitution of this integral into the previous equation leads to the standard expression for LG beams:
\bea
U({\bf x})\!\!&=&\!\!2\pi kP_0\frac{(-1)^{n}n!\left[(N+1)^2-\ell^2\right]^{\frac{2n+1}4}(N+1)^{\ell-\frac12}}{2^{\frac N2}N!(\ell+n)!}\nonumber\\
& &\times\left(\sqrt{1+r}+\sqrt{1-r}\right)^{\ell}\exp\left(\ui\frac{\pi}4-\frac{N+1}2\right)  \label{LGbeam}\\
& &\times\exp\!\left[\!-(N+1)\frac{\rho^2}{Q_0^2}\!\right]\!\exp(\ui \ell\vartheta)\frac{\rho^{\ell}}{Q_0^{\ell}}L_n^{(\ell)}\!\!\left[2(N+1)\frac{\rho^2}{Q_0^2}\right]\!, \nonumber
\eea
where $(\rho,\vartheta)$ are polar coordinates for ${\bf x}$ and $\ell=N-2n$ is the vorticity of the beam. 

\section{Hermite-Gaussian beams}
\label{sec:AppHG}

For HG beams, the Poincar\'e path is a straight line, chosen here in the vertical direction for convenience. Since this line is the projection of a circle of radius $r$ over the unit sphere's surface, we parametrize it as ${\bf s}=(\sqrt{1-r^2},r\sin\eta)$. The angles determining the parametrized point in the Poincar\'e sphere are then
\bea
\theta&=&\arccos\left(\sqrt{1-r^2\cos^2\eta}\right),\\
\phi&=&\arctan\left(\sqrt{1-r^2},r\sin\eta\right).
\eea
From here it is easy to find the medial axes to be parabolas of the form
\be
{\bf t}_{1,2}=\left(\frac{\sqrt{1-r^2}\pm1}2+\frac{\sqrt{1-r^2}\mp1}2\sin^2\eta,r\sin\eta\right),\ee
and the caustics, given by the mapping $\Z({\bf q}_{1,2})=\sqrt{\Z({\bf t}_{1,2})}$, to be straight lines:
\bea
{\bf q}_1&=&\left(\pm\nu_x,\pm\nu_y\sin\eta\right),\\
{\bf q}_2&=&\left(\pm\nu_x\sin\eta,\pm\nu_y\right),
\eea
where
\bea
\nu_x&=&\sqrt{\frac{1+\sqrt{1-r^2}}2}=\sqrt{\frac{2m+1}{2(N+1)}},\\
\nu_y&=&\sqrt{\frac{1-\sqrt{1-r^2}}2}=\sqrt{\frac{2n+1}{2(N+1)}},
\eea
with $m=N-n$.

While the general parametrization of the ray positions employed in this document so far (in which $\tau=0$ corresponds to a vertex of the ellipses) gives the correct result, for this case there is an equivalent but more convenient parametrization given by
\bea
{\bf Q}&=&\left[Q_0\nu_x\cos\left(\tau-\frac{\eta}2\right),Q_0\nu_y\cos\left(\tau+\frac{\eta}2\right)\right],\\
{\bf P}&=&\left[-P_0\nu_x\sin\left(\tau-\frac{\eta}2\right),-P_0\nu_y\sin\left(\tau+\frac{\eta}2\right)\right].
\eea
We can then change variables to $\tau_{x,y}=\tau\pm\eta/2$ (with unit Jacobian), so the position and momentum vectors simplify to
\bea
{\bf Q}&=&\left(Q_0\nu_x\cos\tau_x,Q_0\nu_y\cos\tau_y\right),\\
{\bf P}&=&\left(-P_0\nu_x\sin\tau_x,-P_0\nu_y\sin\tau_y\right),
\eea
and the OPL is now easily found to be
\bea
L&=&\int_0^{\tau_x}{\bf P}\cdot\frac{\partial{\bf Q}}{\partial\tau_x}\ud\tau_x+\int_0^{\tau_y}{\bf P}\cdot\frac{\partial{\bf Q}}{\partial\tau_y}\ud\tau_y\nonumber\\
&=&\frac{Q_0P_0}2\left[\nu_x^2\left(\tau_x-\frac{\sin2\tau_x}2\right)+\nu_y^2\left(\tau_y-\frac{\sin2\tau_y}2\right)\right].
\nonumber \\
\eea
The fact that the total OPL is separable as a sum of a part that depends only on the $x$ components and a part that depends only on the $y$ components heralds the separability of the problem in Cartesian coordinates. 

This separability becomes evident when considering the wave field estimate in Eq.~(9),  
where by assuming $A(\eta)$ to be constant (set to unity for simplicity), one arrives at the exact result, which is of course separable. With $\Gamma=P_0/Q_0$, the Jacobian in Eq.~(9)  
gives
\be
\frac{\partial(\Gamma{\bf Q}+\ui {\bf P})}{\partial(\tau_x,\tau_y)}=P_0^2\frac r2\exp[-\ui(\tau_x+\tau_y)],
\ee
and  Eq.~(9)  
itself reduces to
\bea
U({\bf x})
&=&\frac{kP_0}{2\pi}\sqrt{\frac r2}\,I_x\left(\frac x{Q_0}\right)I_y\left(\frac y{Q_0}\right),
\eea
where each of the separable $x$- and $y$-dependent parts includes an integral over $\tau_x$ or $\tau_y$. The integral for the $x$-dependent part is
\bea
I_x(\bar{x})&=&\int_0^{2\pi}\exp\left[-\ui\frac{\tau_x}2-(N+1)(\bar{x}-\nu_x\cos\tau_x)^2\right]\nonumber\\
& & \times \exp\left[\ui(N+1)\nu_x^2\left(\tau_x-\frac{\sin2\tau_x}2\right)\right]\nonumber\\
& & \times \exp\left[-2\ui(N+1)(\bar{x}-\nu_x\cos\tau_x)\nu_x\sin\tau_x\right]\,\ud\tau_x\nonumber\\
&=&\exp\left[-(N+1)\bar{x}^2-(N+1)\frac{\nu_x^2}2\right]\nonumber\\
& & \times\int_0^{2\pi}\exp\left[\ui\left((N+1)\nu_x^2-\frac12\right)\tau_x\right]\nonumber\\
& & \times \exp\left\{\frac{N+1}2\left[4\bar{x}\nu_x\exp(-\ui\tau_x)-\nu_x^2\exp(-2\ui\tau_x)\right]\right\}\,\ud\tau_x\nonumber\\
&=&\exp\left[-(N+1)\bar{x}^2-\frac{2m+1}4\right]\int_0^{2\pi}\!\!\exp\!\Bigg\{\ui m\tau_x\nonumber\\
& & - \frac{N+1}2\left[\nu_x^2\exp(-2\ui\tau_x)-4\bar{x}\nu_x\exp(-\ui\tau_x)\right]\Bigg\}\,\ud\tau_x.
\eea
By performing the change of variable $u=\nu_x\exp(-\ui\tau_x)$, the integral can be converted into an integral over a circle of radius $\nu_x$ centered at the origin in the complex $u$ plane, which can be easily solved by using residue theory:
\bea
I_x(x)\!\!&=&\!\!-\ui\nu_x^m\exp\left[-(N+1)\bar{x}^2-\frac{2m+1}4\right]\nonumber\\
& & \times\exp\left[2(N+1)\bar{x}^2\right]\oint u^{-m-1}\exp\!\left[-\frac{N+1}2(u-2\bar{x})^2\right]\ud u\nonumber\\
&=&2\pi\frac{\nu_x^m}{m!}\exp\left[-(N+1)\bar{x}^2-\frac{2m+1}4\right]\nonumber\\
& & \times\exp\!\left[2(N+1)\bar{x}^2\right]\lim_{u\to0}\frac{\ud^m}{\ud u^m}\exp\left[-\frac{N+1}2(u-2\bar{x})^2\right]\nonumber\\
&=&2\pi\frac{\nu_x^m}{2^mm!}\exp\left[-(N+1)\bar{x}^2-\frac{2m+1}4\right]\nonumber\\
& &\times(-1)^m\exp\left[2(N+1)\bar{x}^2\right]\frac{\ud^m}{\ud \bar{x}^m}\exp\left[-2(N+1)\bar{x}^2\right]\nonumber\\
&=&\!\!\frac{2\pi\nu_x^m}{2^mm!}\exp\!\left[\!-(N+1)\bar{x}^2\!-\!\frac{2m+1}4\!\right]\!H_{m}\!\!\left[\!\sqrt{2(N+1)}\bar{x}\!\right],
\eea
where $H_m$ is a Hermite polynomial of the first kind. The integral $I_y$ is identical, with $\bar{y}$ and $n$ replacing $\bar{x}$ and $m$, respectively. The field is then the standard expression for HG beams, namely
\bea
U({\bf x})&=&2\pi kP_0\frac{(2m+1)^{\frac{2m+1}4}(2n+1)^{\frac{2n+1}4}}{2^\frac{3N+1}2(N+1)^\frac{N+1}2m!n!}\exp\left[-\frac{N+1}2\right] \nonumber\\
&\times&\exp\left[-(N+1)\frac{x^2+y^2}{Q_0^2}\right]\nonumber\\
&\times&H_{m}\!\!\left[\sqrt{2(N+1)}\frac x{Q_0}\right]H_n\!\!\left[\sqrt{2(N+1)}\frac y{Q_0}\right].
\label{HGbeam}
\eea

\end{document}